\documentclass[]{raa}            
\usepackage{graphicx,times}
\usepackage{natbib}
\usepackage{amssymb,amsmath}

\providecommand{\bm}[1]{\mbox{\boldmath$#1$\unboldmath}}

\begin{document}

   \title{Magnetic cycles of Sun-like stars with different levels of coronal
and chromospheric activity -- comparison with the Sun}
 \volnopage{ {\bf 2016} Vol.\ {\bf 00} No. {\bf }, 000--000}
   \setcounter{page}{1}

   \author{E. A. Bruevich
   \inst{},
   \ V. V. Bruevich
   \inst{}
   \and E. V. Shimanovskaya
   \inst{}
   }

 \institute{Sternberg Astronomical Institute, Moscow State
 University, Universitetsky pr., 13, Moscow 119992, Russia;
            {\it red-field@yandex.ru, brouev@sai.msu.ru, eshim@sai.msu.ru,}\\
   \vs \no
   {\small Received [year] [month] [day]; accepted [year] [month] [day] }
}

\abstract{The atmospheric activity of the Sun and Sun-like stars is
analyzed involving observations from HK-project at the Mount Wilson
Observatory, the California and Carnegie Planet Search Program at
the Keck and Lick Observatories and the Magellan Planet Search
Program at the Las Campanas Observatory. We show that for stars of
F, G and K spectral classes, the cyclic activity, similar to the
11-yr solar cycles, is different: it becomes more prominent in
K-stars. Comparative study of Sun-like stars with different levels
of the chromospheric and coronal activity confirms that the Sun
belongs to stars with the low level of the chromospheric activity
and stands apart among these stars by the minimum level of its
coronal radiation and the minimum level of its variations of the photospheric flux.}

 \authorrunning{ E. A. Bruevich, V. V. Bruevich and E.V. Shimanovskaya}            
   \titlerunning{Magnetic cycles of Sun-like stars }  
   \maketitle

\section{Introduction}
{\label{S:intro}}

The study of the magnetic activity of the Sun and Sun-like stars is of
fundamental importance for astrophysics. This activity of stars
leads to the complex of composite electromagnetic and hydrodynamic
processes in their atmospheres. Local active regions, which are
characterized by a higher value of intensity of the local magnetic
field, are: plagues  and spots in photospheres, CaII flocculae in
chromospheres and prominences in coronas.

More powerful coronas are possessed by stars displaying irregular
variations of their chromospheric emission, while stars with cyclic
activity are characterized by comparatively modest X-ray
luminosities and ratios of the X-ray to bolometric luminosity
$L_X/L_{Bol}$. This indicates that the nature of processes
associated with the magnetic-field amplification in the convective
envelope changes appreciably in the transition from small to large
dynamo numbers, directly affecting the character of the
($\alpha-\Omega$) dynamo. Due to the strong dependence of both the
dynamo number and the Rossby number on the speed of axial rotation,
earlier correlations were found between various activity parameters and
the Rossby number in Bruevich et al. (2001).

It is difficult to predict the evolution of each active region in
details. However, it has long been established that the total change
of active areas integrated over the entire solar or stellar disk is
cyclical not only in solar activity but in stellar activity too (see
Baliunas et al. 1995; Kollath \& Olah 2009; Morgenthaler et al.
2011; Bruevich \& Kononovich 2011).

It is well known that the duration of the 11-yr cycle of solar
activity (Schwabe cycle) ranges from 7 to 17 years according to a
century and a half of direct solar observations.

The most sensitive indicator of the chromospheric activity (CA) is the
Mount Wilson $S$ - index ($S_{HK}$) - the ratio of the core of the
CaII H\&K lines to the nearby continuum (Vaugan \& Preston 1980).
Now the CaII H\&K emission was established as the main indicator of
CA in lower main sequence stars.

The HK-project of Mount Wilson observatory is one of the first and
still the most outstanding program of observations of Sun-like stars
(see Baliunas et al. 1995; Lockwood et al. 2007). One of the most
important results of the HK-project was the discovery of "11-yr"
cycles of activity in Sun-like stars. Durations of CA cycles, found
for 50 different stars of late spectral classes (F, G and K), vary
from 7 to 20 years according to HK-project observations.

Currently, there are several databases that include thousands of
stars with measured fluxes in the chromospheric lines of CaII H\&K
emission cores (see Wright et al. 2004; Isaacson \& Fisher 2010;
Arriagada 2011; Garcia et al. 2010; Garcia et al. 2014). However,
only for a few tens of stars the periods of magnetic activity cycles
are known (Baliunas et al. 1995; Radick et al. 1998; Lockwood et al.
2007; Olah et al. 2009; Morgenthaler et al. 2011).

In our work, we consider the following databases of observations of
Sun-like stars with known values of $S_{HK}$:

 1. HK-project -- the program in which the Mount Wilson $S$ value was first
defined. Later the Mount Wilson "$S$ value" ($S_{HK}$) became the
standard metric of CA -- the basic value with which all future
projects of stellar CA observations are compared and calibrated.

2. The California \& Carnegie Planet Search Program which includes
observations of approximately 1000 stars at Keck \& Lick
observatories  in chromospheric CaII H\&K emission cores.
$S_{HK}$ indexes of these stars are converted to the Mount Wilson system
 (Wright et al. 2004). From these measurements, median activity
levels, stellar ages, and rotation periods from general
parameterizations have been calculated for 1228 stars, $\sim$ 1000
of which have no previously published $S$-values.

3. The Magellan Planet Search Program which includes Las Campanas
Observatory CA measurements of 670 F, G, K and M
main sequence stars of the Southern Hemisphere. $S_{HK}$-indexes of these
stars are also converted to the Mount Wilson system (Arriagada
2011).

We believe that the stars in these databases are best described the
stars which are similar to the Sun in mass and evolutionary state.
They are main-sequence stars with a $(B-V)$ color between 0.48 and
1.00 (the Sun has a $(B-V)$ color of 0.66). Alternatively, a
definition based on spectral type can be used, such as F5V through
K5V.

The aims of our paper are: 1) a study of the place of the Sun
among stars with different levels of chromospheric and coronal
activity belonging to the main sequence on the Hertzsprung-Russell
diagram; 2) a comparative analysis of chromospheric, coronal and
cyclic activity of the Sun and Sun-like stars of F, G and K spectral
classes.

\begin{figure}[h!!!]
   \centering
   \includegraphics[width=11.0cm, angle=0]{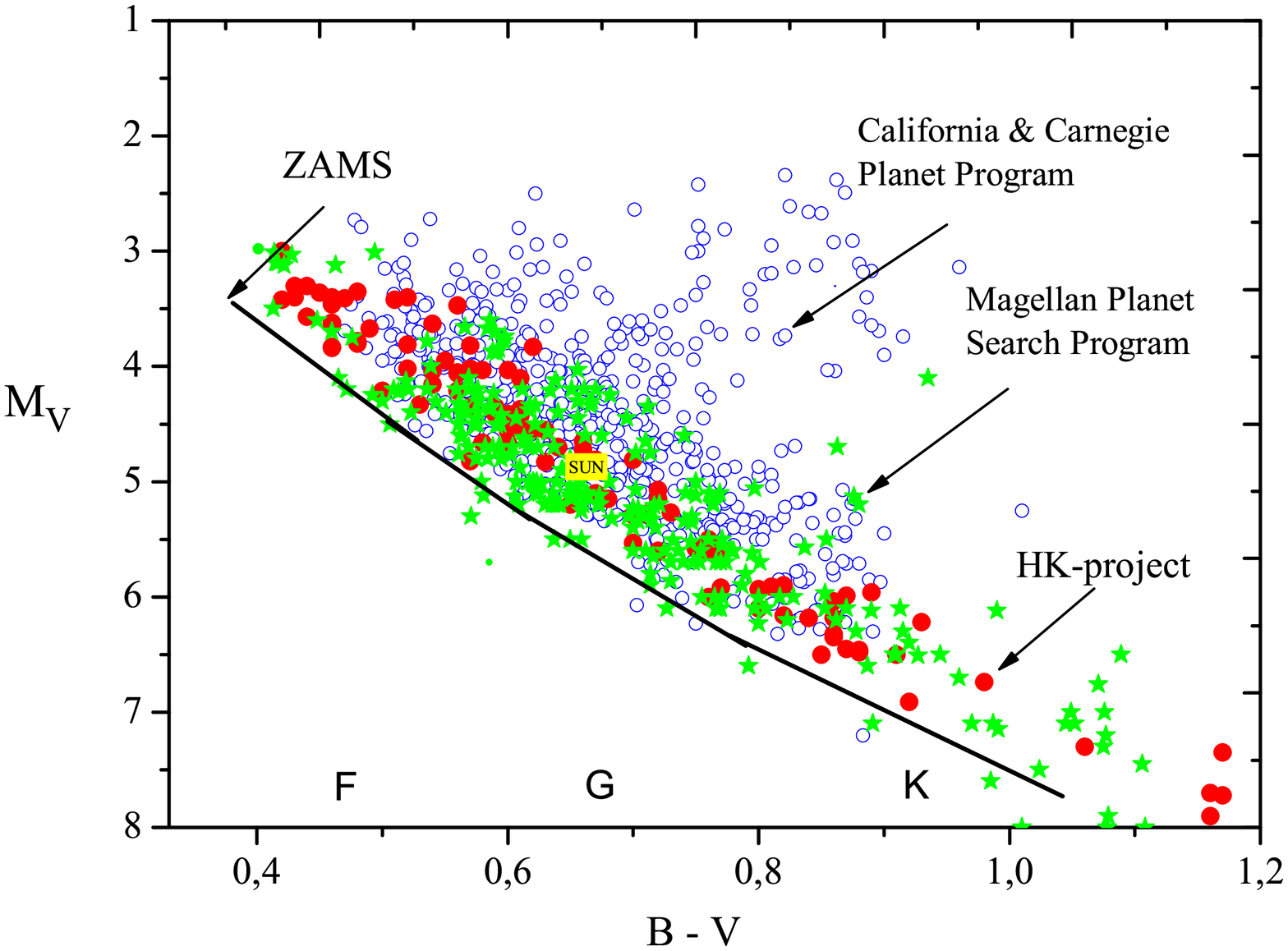}

   \begin{minipage}[]{100mm}
\parbox[t]{120mm}{Fig. 1. The Sun among the Sun-like stars from different observational
 Programs on the Hertzsprung-Russell diagram. }

\end{minipage}
   \end{figure}

\section{The place of the Sun among the Sun-like stars with different levels of chromospheric and coronal activity}
{\label{S:place}}

In Figure 1, the Sun and stars from different observed samples are
presented on the Color-Magnitude Hertzsprung-Russell diagram. There
are  1000 Sun-like stars from Wright et al. (2004), observed in the 
Program of Planet Search (open circles);  660 Sun-like stars from
Arriagada (2011) observed in the Magellan Planet Search Program (asterisks) and 110 Sun-like
stars and the Sun from Baliunas et al. (1995), observed in the Mount Wilson HK-project (filled circles). The
solid line represents the Zero Age Main Sequence (ZAMS) on the
Hertzsprung-Russell diagram.

Stars, which are close to the ZAMS in Figure 1, have the lowest age
among all other stars: log(Age/yr) is about 8 - 8.5. The older the
star is, the farther it is from the ZAMS. Ages of all the stars in 
Figure 1 varies from $10^8$ to $10^{10}$ years.

We can also see that some stars significantly differ from the Sun by
the absolute magnitude $M_V$ and color indices $(B-V)$.

In different samples of stars from  Planet Search Programs, some
observers included a number of subgiants with larger values of
magnitudes $M_V$ among  Sun-like stars belonging to the
main sequence. We try to exclude them from our analysis.

In Figure 2, one can see that stars have significantly different
values of the  $S_{HK}$-index, which determines their CA. There are stars
from the Program of Planet Search (open circles), the Magellan Planet Search
Program (asterisks) and 110 stars and the Sun from the HK-project
(filled circles).

We cannot see the close relationship of CA of stars from our samples
versus the color index in Figure 2.

Figures 1,2 show that the more sun-like stars in pattern, the greater the spread in their luminosities $M_V$ and
chromospheric activity CA versus the Sun value.

It is important to note that the study of CA of  stars are performed for 
very large data sets. Zhao et al. 2015 have studied a sample of
$\sim$ 120 000 F, G and K stars from the LAMOST DR1 archive, which
is an unprecedentedly large sample for a study of the H\&K CaII emission.
The $ \delta S$ index was measured for these stars, calculated as
the difference between standard $S$-index and a 'zero' emission line
fitted using several of the least active stars across the whole
range of $T_{eff}$. It was shown that active stars lie closer to the
Galactic plane but inactive stars tend to be farther away from the
Galactic plane.

For another large statistical sample -- 2600 stars of the California
Planet Search Program -- the lower envelope of CA level $S_{BL}$
($S_{HK}$ of the Basic Level) was defined as a function (polynomial fit)
of $B-V$ for main sequence stars over the color range $0.4<B-V<1.6$,
see Isaacson et al. (2010).

We show the $S_{BL}$ dependence in Figure 2. It is seen that the
Basic CA Level $S_{BL}$  begins to rise when $B-V>1$. Isaacson et
al. (2010) believed that this increase of $S_{BL}$ is due to
decrease of continuum flux for redder stars: $S_{HK}$ is defined as
the ratio of H\&K CaII emission to the nearby continuum.

According to Isaacson et al. (2010), mean levels of CA
(corresponding to the uniform Mount Wilson $S$-index) for stars of
the spectral class F are higher than that of the G-stars. On the
other hand, for stars of K and M spectral classes, mean levels of CA
are also higher than that of G-stars. However it can be noted that
the level of CA of the Sun is slightly below than the average level
of CA of stars belonging to the main sequence.

The $S_{HK}$ is affected by line blanketing in the continuum regions
that increases with (B-V) color index. This effect gives mistakes in
the comparison of CaII activity for stars of different color. 

For describing the CA, it will be more correct to use the index $log
R'_{HK}$. The parameter $~log R'_{HK}$ is calculated from the mean
$S_{HK}$ and the (B-V) color. It was originally formulated by Noyes et
al. (1984). The HK-project stars sample ranges from $log R'_{HK}$= - 4.4
(young stars) to -5.3 (old stars). The solar value of $log R'_{HK}$=
- 4.94.

In our paper we use $S_{HK}$ when we study the observed time series
("light curves" of the H\&K CaII emission) but below in this paper we
will use $log R'_{HK}$ as the more correct index of the the CA.

\begin{figure}[h!!!]
   \centering
   \includegraphics[width=11.0cm, angle=0]{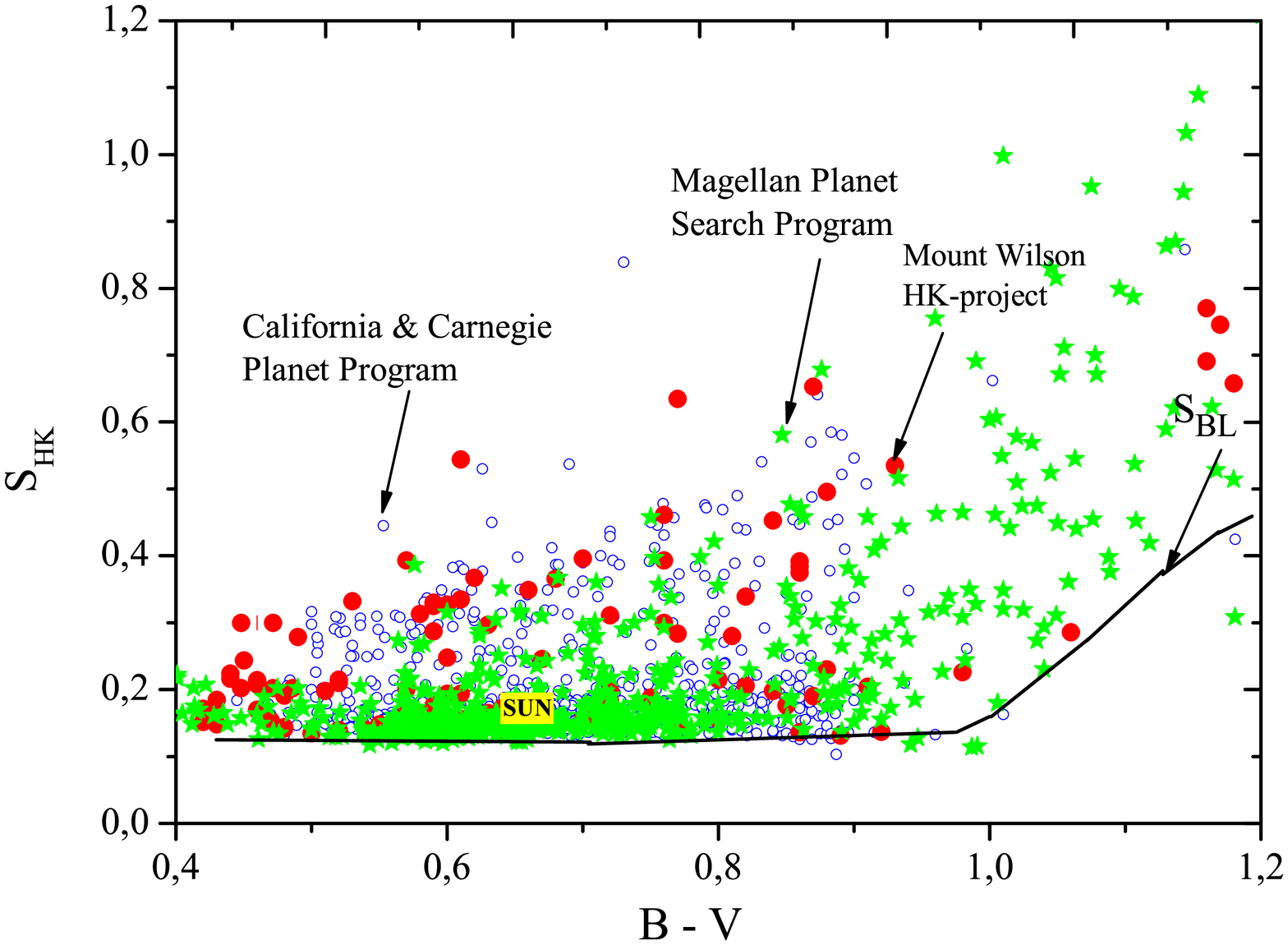}

   \begin{minipage}[]{100mm}
\parbox[t]{120mm}{Fig. 2. Chromospheric activity of F, G, K and M stars from three observational Programs. }

\end{minipage}
   \end{figure}

In Sun-like stars of late spectral classes, X-rays are generated by
the magnetically confined plasma known as the corona (see Vaiana et
al. 1981), which is heated by the stellar magnetic dynamo. The
observed decrease of the X-ray emission between pre-main sequence
young stars and older stars can be attributed to the rotational
spin-down of a star, driven by mass loss through a magnetized
stellar wind (Skumanich 1972).

The magnetic fields of the Sun and solar-type stars are believed to
be generated by flows of conductive matter. Such processes are
commonly called a hydromagnetic dynamo or simply a dynamo.

The main parameter of solar and stellar dynamo models is the dynamo
number $ D = \alpha \Omega R^3/ \eta^2 $,  where $\Omega$ is the
rotation rate of a star, R is its radius, $\eta$ is the effective
magnetic diffusivity, and the parameter $\alpha$ defines the
intensity of field generation by the cyclonic convection (see Parker
1955; Vainshtein et al. 1980). The generation of magnetic fields is
ultimately governed by hydrodynamic processes in stellar convection
zones. The Sun's radiative interior is surrounded by a convection
zone that occupies the outer $30 \%$ by radius; in most part of this
region the angular velocity $\Omega$ is constant on conical surfaces
and there is a balance between Coriolis and baroclinic effects. At
the base of the convection zone there is an abrupt transition in the
tachocline to the almost uniformly rotating radiative zone.
Superimposed on this pattern are zonal shear flows (torsional
oscillations) that vary with the solar cycle, appearing as a branch
of enhanced rotational velocity that coincides with the activity
belt on the Sun and solar-type stars. As expected, the torsional
oscillations on the Sun have an 11-year period, a half of the
underlying magnetic 22-year cycle. The main implications of dynamo
theory, as applied to stars, has revealed the main features of the cyclic
behavior in the Sun and stars, but it involves ad hoc
assumptions, that helps us to compare the dynamo theory results with
observations.

The nonlinear dynamo model allows to explain the phenomenon of the
irregular appearance of the periods of Grand minima along with the
regular cycles of activity. In the nonlinear dynamo theory there is an
important parameter -- the critical dynamo number $D_C$ for the field
generation. The magnetic field decays for D below Its critical value
namely, when $D < D_C$ and rises with time for $D > D_C  $. In a
certain range of dynamo numbers, two types of solutions are
possible: decaying oscillations of weak fields (known as Maunder
minimum in solar observations) and magnetic cycles with a constant
and large amplitude (known as the standard 11-year cycles), see
Kitchatinov \& Olemskoy (2010).

The rotation rate of the Sun and solar-type stars decreases with time (Skumanich, 1972), 
with a
feedback existing between the dynamo and rotation: the higher the magnetic activity, is the larger the rate of angular momentum loss. The reducing of the spin rate  may ultimately bring the stars to the threshold of large-scale dynamo action. The states
of low activity are actually observed only in old stars(Wright et al., 2004). If the proposed picture is valid, then observations can reveal a sharp decrease in reducing of the spin rates
 for old stars that exhibit grand minima of activity.

\begin{figure}[h!!!]
   \centering
   \includegraphics[width=11.0cm, angle=0]{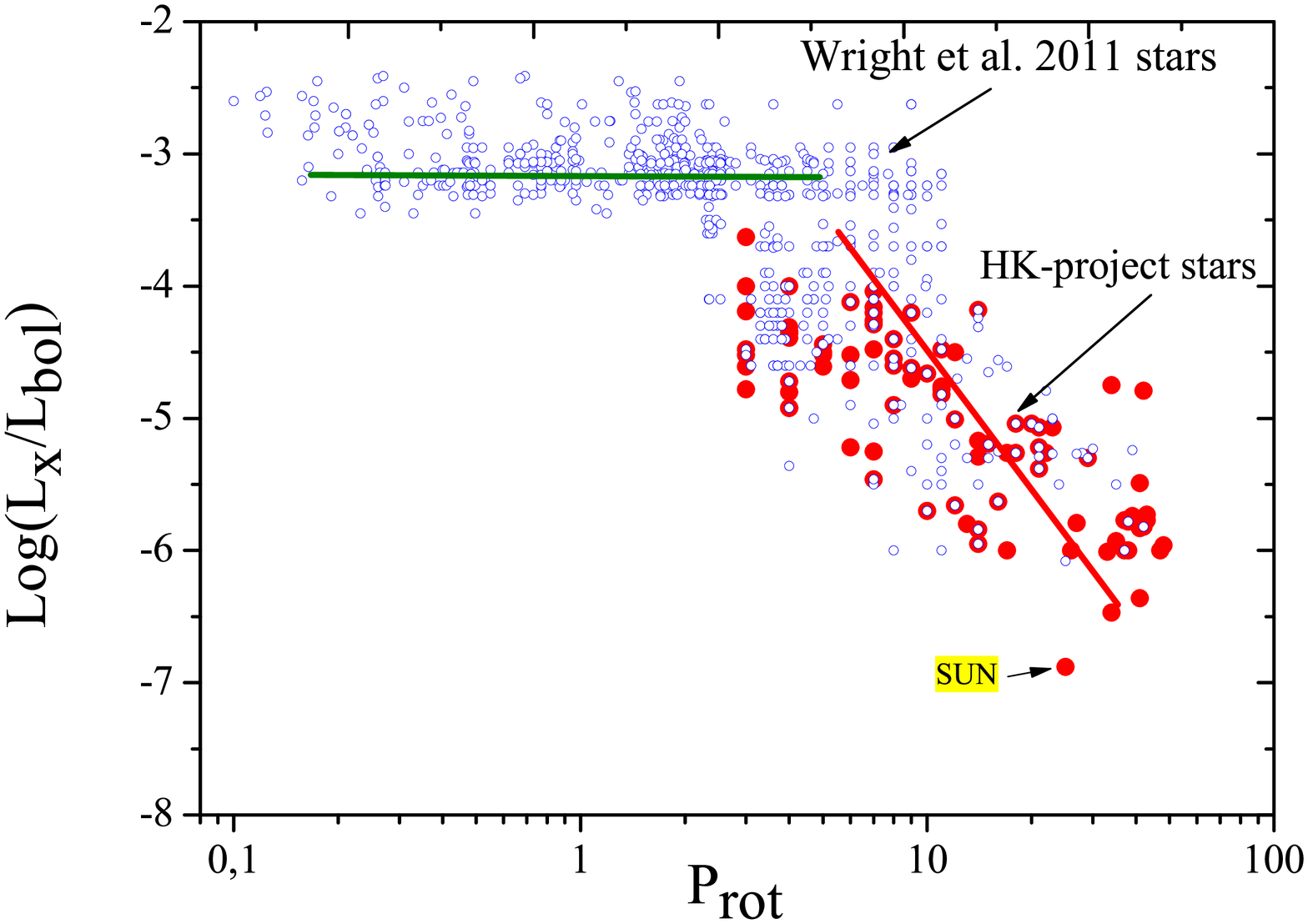}

   \begin{minipage}[]{100mm}
\parbox[t]{120mm}{Fig. 3. X-ray to bolometric luminosity ratio plotted against rotation period. }

\end{minipage}
   \end{figure}

A relationship between the stellar rotation and the X-ray luminosity $L_X$
 was first described by Pallavicini et al. (1981).
 In Figure 3 we show the $log(L_X/L_{bol})$ versus $P_{rot}$ for
 stars from the catalogue of stars, the details of which are
 presented in Wright et al. (2011)- open circles, and for the stars of the HK-project
 presented in Baliunas et al. (1995) - filled circles.
In Write et al. (2011) there was done the convertation of all X-ray
luminosities of the stars from the catalogue of 824 stars to the ROSAT
0.1-2.4 keV energy band. The selected data of X-ray luminosities
$L_X/L_{bol}$ of 80 HK-project stars was also taken from the ROSAT
All-Sky Survey (Bruevich et al. 2001). This sample also includes the
Sun using the values of $log L_X/L_{bol}=-6.88$ (for medium level of
activity) and $P_{rot}=26.09$.
 Figure 3 demonstrates that there are two main regimes of coronal activity: a linear regime
 where activity increases with decreasing the rotation period, and a saturated
 regime where the X-rays luminosity ratio is constant with
 $log(L_X/L_{bol})=~-3.13$ (see Wright et al. 2011, Wright et al. 2013).

 We can see that the HK-project stars
 which are not enough young and active, belong to stars of linear
 regime in Figure 3. We can also note that the Sun confirms its place among
 Sun-like stars: it has relatively low
 level of X-rays luminosity among all the stars in the studied sample.

Below in Figure 7b we also show the place of the solar coronal
activity among the coronal activity of stars of the HK-project. It
can be noted that the Sun is at the place with absolutely
lowest level of coronal activity among Sun-like HK-project stars.

One can also note that solar photometric radiation changes very
little in the activity cycle, less than 0.1 \%. The simultaneous monitoring of photometric and chromospheric H\&K CaII
 emission data series of stars similar to the Sun in age and average activity
level showed that there is an empirical correlation between the
average stellar CA level and the photometric variability. In
general, more active stars show larger photometric variability. The
Sun is significantly less variable that indicates by the empirical
relationship, see Shapiro et al. (2013). It was found that on a long
time scale the position of the Sun on the diagram of photometric
variability versus CA changers is not constant in time. So Shapiro
et al. (2013) suggested that the temporal mean solar variability
might be in agreement with stellar data.

But at present we can see that the Sun confirms its unique place
among Sun-like stars: its photometric variability is unusually
small. The observational verification has confirmed in Lockwood et al. (2007): the Lowell observatory photometric
observations  of  33 stars of the HK-project
revealed the fact that the
 photometric variability of the Sun during the cycle of magnetic activity
 is much less than the photometric variability of other HK-project
 stars.

\vskip12pt
\section{Observations of HK-project stars}
\vskip12pt

It can be noted that among the databases of observations of Sun-like
stars with known values of $S_{HK}$ the sample of stars of the 
HK-project was selected most carefully in order to study stars
which are analogues of the Sun.
Moreover, unlike different Planet Search Programs of observations of
Sun-like stars, the Mount Wilson Program was specifically developed
for a study of a Sun-like cyclical activity of main sequence F,
G and K-stars (single) which are the closest to the "young Sun" and
"old Sun".

The duration of observations (more than 40 years) in the
HK-project has allowed to detect and explore the cyclical activity
of the stars, similar to 11-yr cyclical activity of the Sun. First
O. Wilson began this program in 1965. He attached great importance
to the long-standing systematic observations of cycles in the stars.
Fluxes in passbands of 0.1 nm wide and centered on the CaII H\&K emission cores
have been monitored in 111 stars of the spectral
type F2-K5 on or near main sequence on the Hertzsprung-Russell
diagram (see Baliunas et al. 1995; Radick et al. 1998; Lockwood
et al. 2007).

For the HK-project, stars were carefully chosen according to those
physical parameters, which are  most close to the Sun: cold, single
stars -- dwarfs, belonging to the main sequence. Close binary systems
are excluded.

Results of  joint observations of the HK-project  radiation
fluxes and periods of rotation gave the opportunity for the first
time in stellar astrophysics (Noyes et al. 1984) to detect the rotational modulation of
the observed fluxes. This meant that on the
surface of a star there are inhomogeneities those are living and
evolving in several periods of rotation of the stars around its
axis. In addition, the evolution of the periods of rotation of the
stars in time clearly pointed to the fact of existence of the star's
differential rotations similar to the Sun's differential rotations.

The authors of the HK-project with use of frequency analysis of
the 40-year observations have discovered that
 the periods of 11-yr cyclic activity vary
little in size for the same star (Baliunas et al. 1995); Lockwood et
al. 2007). Durations of cycles vary from 7 to 20 years for
different stars. It was shown that stars with cycles  represent
about 30 \% of the total number of studied stars.

\vskip12pt
\section{Cyclic activity of HK-project stars. From periodogram to wavelet analysis.}
\vskip12pt

\begin{figure}[h!!!]
   \centering
   \includegraphics[width=11.0cm, angle=0]{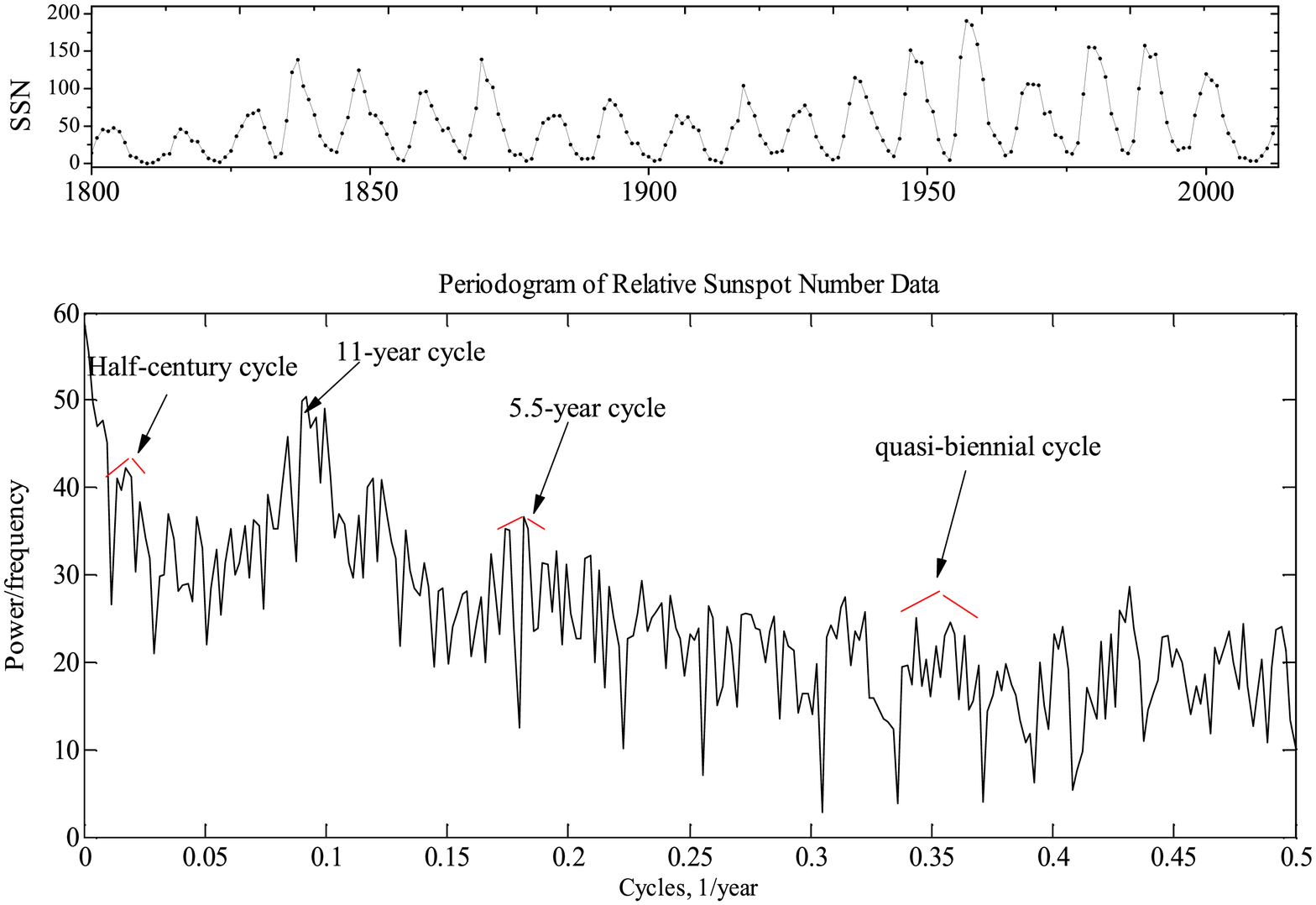}

   \begin{minipage}[]{100mm}
\parbox[t]{120mm}{Fig. 4. The periodogram of the yearly averaged relative sunspot
numbers for 1800 -- 2013 observations. }

\end{minipage}
   \end{figure}

The evolution of active regions on a star on a time scale of about
10 years determines the cyclic activity similar to the Sun.

For 111 HK-project stars the, periodograms were computed for each
stellar record in order to search for activity cycles (Baliunas
et al. 1995). The significance of the height of the tallest peak of
the periodogram was estimated by the false alarm probability (FAP)
function (see Scargle 1982). The stars with cycles  were classified
as follows: if for the calculated $P_{cyc} \pm \Delta P$ the FAP
function $\leqslant 10^{-9}$ then this star is of the "Excellent" class
($P_{cyc} $ is the period of the cycle). If $ 10^{-9} \leqslant FAP
\leqslant 10^{-5} $ then this star is of the "Good" class. If $ 10^{-5}
\leqslant FAP \leqslant 10^{-2} $ then this star is of the "Fair" class.
If $ 10^{-2} \leqslant FAP \leqslant 10^{-1} $ then this star is of the
"Poor" class.

In Baliunas et al. (1995); Radick et al. (1998); Lockwood et al.
(2007) the regular chromospheric cyclical activity of HK-project
Sun-like stars were studied through the analysis of the power
spectral density with the Scargle's periodogram method (Scargle
1982). It was pointed out that the detection of a periodic signal
hidden by a noise is frequently a goal in astronomical data
analysis. So, in Baliunas et al. (1995) the periods of HK-project
stars activity cycles similar to 11-yr solar activity cycle were
determined. The significance of the height of the tallest peak of
the periodogram was estimated by the false alarm probability (FAP)
function. 
Among the 50 stars with detected cycles, only 13
stars (with the Sun), which are characterized by the cyclic activity
of the "Excellent" class, have been found.

We have illustrated the method of cyclic period calculation with
Scargle's periodogram technique on the example of the Sun. We
obtained the periodogram of the  yearly averaged relative sunspot
number or SSN -- Solar Sunspot Number for observations conducted in
1800 -- 2013.

In Figure 4 we present the relative sunspot number yearly averaged
data set (top panel). It is known that direct observations of
sunspots were made only since 1850, and from 1700 to 1850 the
sunspots data have been taken from indirect estimates. This fact, as we
will see below in Figure 5a, affects the quality of time-frequency
analysis for 1700 - 1850 data.

In Figure 4 we show the Scargle's periodogram of the relative
sunspot number for 1800 - 2013 data set (presented in bottom panel).
Our sample periodogram in Fig. 4 shows that the main period of
cycles is equal approximately to 11 years. The level of the
Power/Frequency or Power Spectral Density (PSD) as function of
frequency (1/year) for these observational data shows that the PSD
value for the 11-year cycle is 2 times grater than for nearby PSD
values. The fact that the peak of the 11-year periodicity is not
very sharp shows that the period of the 11-year cycle is not
constant: it changes (for 2 centuries of observations) from 
10 to 12 years.  In Figure 4 we marked the peaks corresponding to
the periods of the second order of smallness: a half-century cycle
phenomena which were first discovered by Vitinsky et al. (1986), 5-5
-year and quasi-biennial cycles. The half-century cycle and 5-5
peaks are not much pronounced (some of percent) against the
neighboring signals. The quasi-biennial peak (due to the fact that
he's very weak on power and is blurred due to the evolution of the
period from 3.5 to 2 years within a single 11-year cycle) is in
general at the noise level. As one can see, the very obvious
peak of PSD, which corresponds to the 11-year cycle, has a value
only about 2 times greater than the surrounding background.

\begin{figure}[h!!!]
\includegraphics[width=70mm]{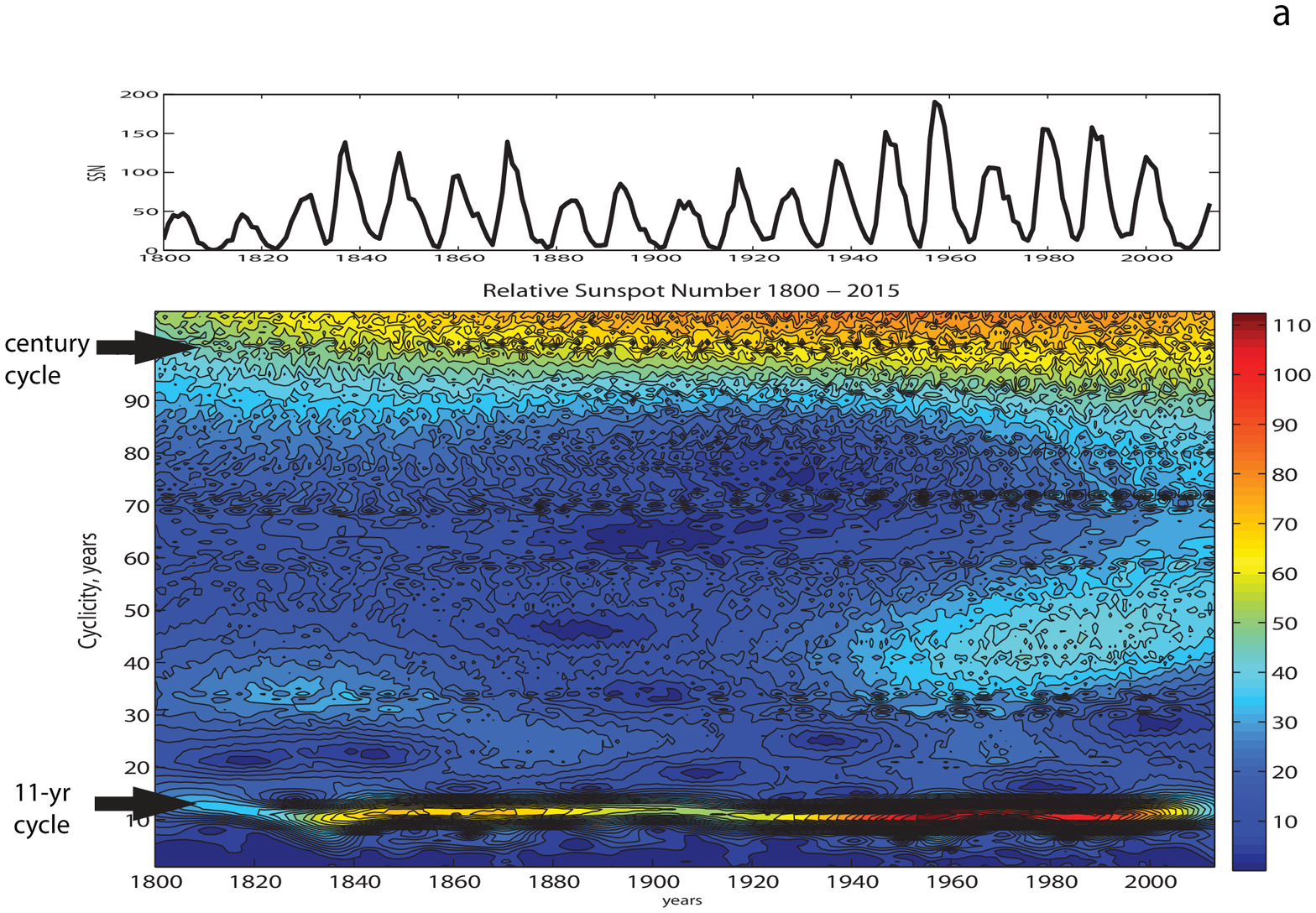}
\includegraphics[width=71mm]{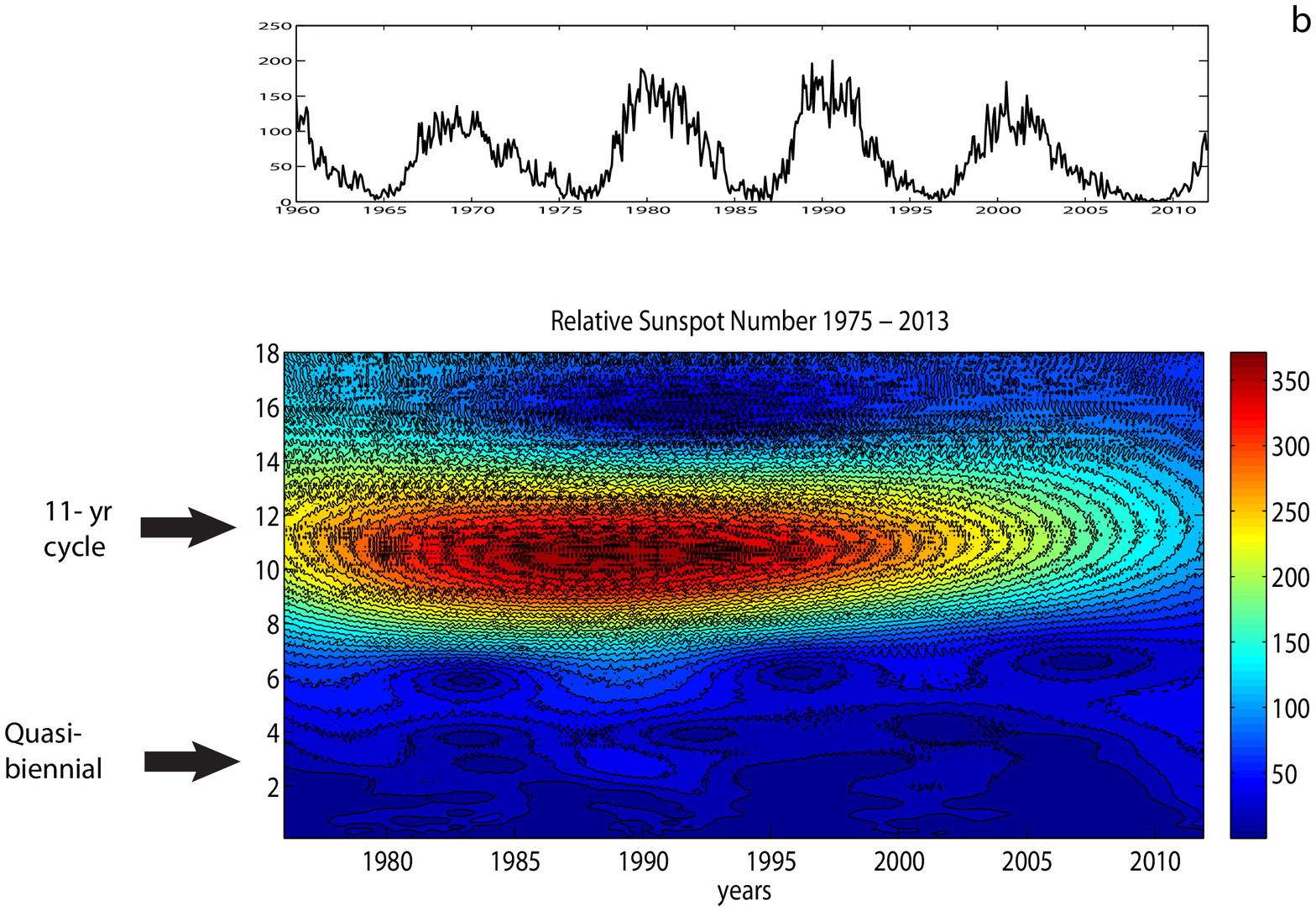}

\end{figure}

\begin{figure}[h!!!]
   \centering
   \includegraphics[width=8.7cm, angle=0]{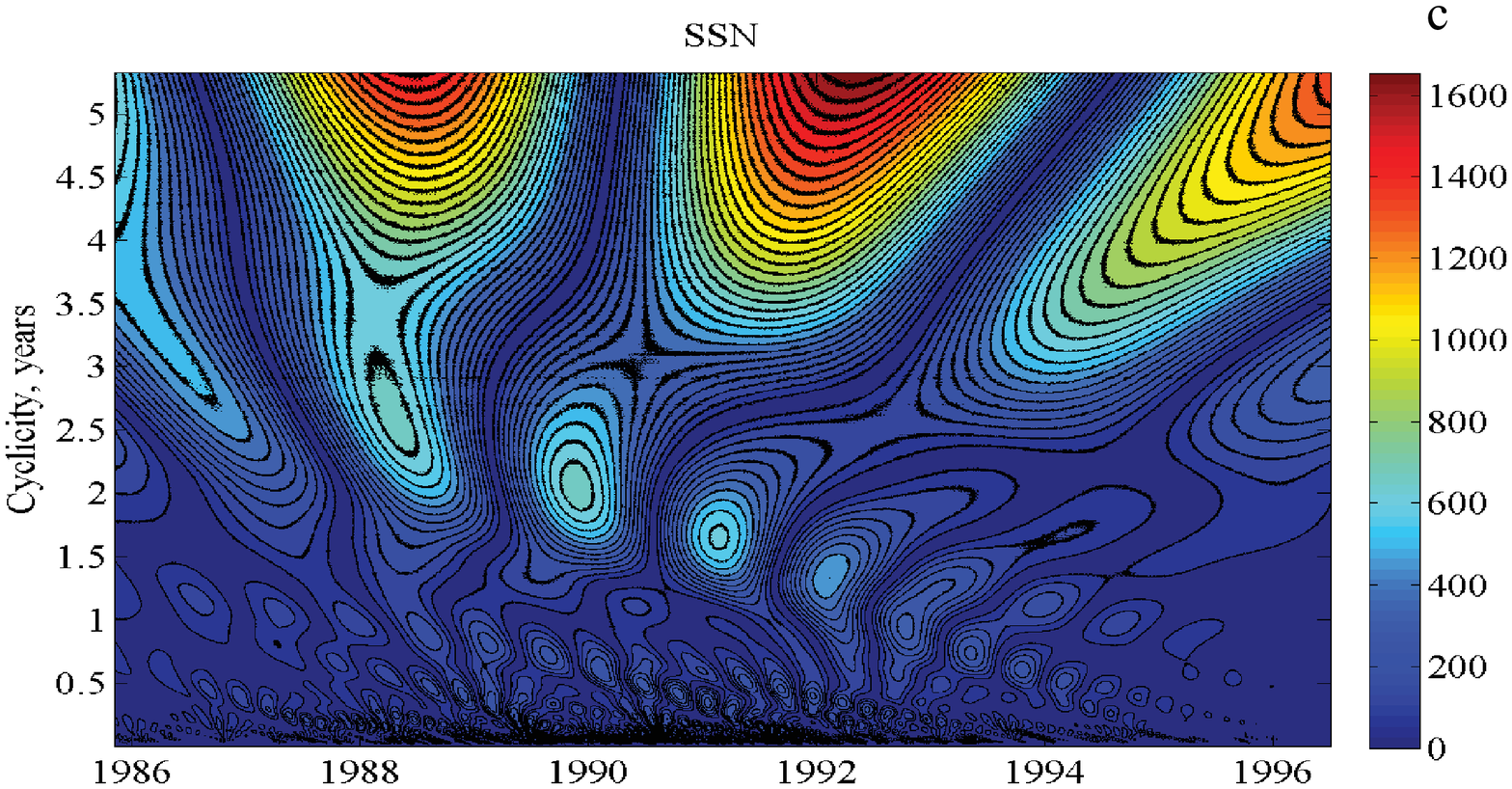}

   \begin{minipage}[]{120mm}
\parbox[t]{120mm}{Fig. 5. Wavelet analysis  of the relative sunspot numbers: (a)
yearly averaged observations from 1700 to 2012; (b) monthly averaged
observations from 1950 to 2012;  (c) daily observations in 22-nd
solar activity cycle. }

\end{minipage}
   \end{figure}

We can see that, unfortunately, this method allows us only to define
a fixed set of main frequencies (that determines the presence of
significant periodicities in the series of observations). In the
case where values of periods change significantly during the
interval of observations, the accuracy of determination of periods
becomes worse. It is also impossible to obtain information about the
evolution of the periodicity in time.

In Kollath \& Olah (2009), different
methods, such as short-term Fourier transform, wavelet, and
generalized time-frequency distributions, have been tested and used for analyzing temporal
variations in timescales of long-term observational data which have
information on the magnetic cycles of active stars and that of the
Sun. It was shown that the application of the wavelet analysis is
preferable when studying a series of observations of the Sun and
stars. Their time-frequency analysis of  multi-decadal variability
of the solar Schwabe (11-yr) and Gleissberg (century)  cycles during
the last 250 years showed that one cycle (Schwabe) varies between
limits, while the longer one (Gleissberg) continually increases. By
analogy from the analysis of the longer solar record, the presence
of a long-term trend may suggest an increasing or decreasing of a
multi-decadal cycle that is presently unresolved in stellar
records of short duration.

A wavelet technique has become popular as a tool for extracting a local
time-frequency information. The wavelet transform differs from the
traditional time-frequency analysis (Fourier analysis, Scargle's
periodogram method) because of its efficient ability to detect and
quantify multi-scale, non-stationary processes (Frick et al. 1997).
The wavelet transform maps a one-dimensional time series $f(t)$ into
the two-dimensional plane, related to time and frequency
scales. Wavelets are the localized functions which are
constructed based on one so-called mother wavelet $\psi(t)$. The
choice of wavelet is dictated by signal or image characteristics
and the nature of the application. Understanding properties of
the wavelet analysis and synthesis, you can choose a mother wavelet
function that is optimized for your application. Thus, we choose a
complex Morlet wavelet which depends on two parameters: a bandwidth
parameter and a wavelet center frequency. The Morlet wavelet allows
one to get results with better spectral resolution. The Morlet
mother wavelet function can be represented as
$\psi(t)=e^{-t^2/a^2}e^{i2\pi t}$.

The standard algorithm of wavelet analysis as applied to
astronomical observations of the Sun and stars was discussed in
detail in Frick et al. (1997), Kollath \& Olah (2009). The choice of the
Morlet wavelet as the best mother wavelet $\psi(t)$ for astronomical
data processing is discussed in Bruevich et al. (2014) on the basis
of comparative analysis of results obtained using different mother
wavelets.

The results of our solar data wavelet analysis are presented on the
time-frequency plane (Figures 5a,b,c). The notable crowding of
horizontal lines on the time-frequency plane around the specific
frequency indicates that the probability of existence of stable
cycles is higher for that frequency (cycle's duration) in accordance
with the gradient bar to the right.

In Figure 5 the results of the continuous wavelet transform
analysis (with help of Morlet mother wavelet) of time series of SSN are presented:
Fig 5a corresponds to yearly averaged data, Fig 5b corresponds to monthly
averaged data, Fig 5c corresponds to daily data. The (X, Y) plane is the
time-frequency plane of calculated wavelet-coefficients C(a, b):
a-parameter corresponds to Y plane (Cyclicity, years), b-parameter
corresponds to X plane (Time, years). The modules of C(a,b)
coefficients, characterizing the probability amplitude of regular
cyclic component localization exactly at the point (a, b), are laid
along the Z axis. In Figure 5 we see the projection of C(a,b) to (a,
b) or (X, Y) plane. This projection on the plane (a, b) with
isolines allows to trace changes of the coefficients on various
scales in time and reveal a picture of local extrema of these
surfaces. It is the so-called skeleton of the structure of the
analyzed process. We can also note that the configuration of the Morlet
wavelet is very compact in frequency, which allows us to determine the
localization of instantaneous frequency of observed signal most
accurately (compared to other mother wavelets).

Figure 5a presents the results of wavelet analysis of the same
yearly averaged sunspot number data set as in Figure 4. The cycles
with century, half-century and 11-yr durations  are marked with
arrows. In Figure 5b we show the results of wavelet analysis of the
monthly averaged sunspot number data set. The cycles with 11-yr and
quasi-biennial durations  are also marked with arrows. Note that for
the study of cycles on a shorter time scale (Figure 5b) we use more
detailed observations of the Sun.

Figure 5a confirms the known fact that the period of the main solar
activity cycle is about 11-yr in the XIX century and is about 10 yr
in the XX century. It is also known that the abnormally long 23-rd
cycle of solar activity ended in 2009 and lasted about 12.5 years.
We can see all this facts in Figures 5a,b. Thus, it can be argued
that the value of a period of the main cycle of solar activity for
past 200 years is not constant and varies by 15-20 \%.

In Figure 5c we show the results of wavelet analysis of daily SSN
data in the solar cycle 22. We analysed that data on the time scale
which is equal to several years and
identified the second order periodicity such as 5.5 years and
quasi-biennial as well as their temporal evolution.

In Olah et al. (2009) a study of time variations of cycles
of 20 active stars based on decades (long photometric or
spectroscopic observations) with a method of time-frequency analysis
was done. They found that cycles of sun-like stars show
systematic changes. The same phenomenon can be observed for the cycles of
the Sun.

Olah et al. (2009) found that fifteen stars definitely show multiple
cycles, the records of the rest are too short to verify a timescale
for a second cycle. For 6 HK-project stars (HD 131156A, HD 131156B,
HD 100180, HD 201092, HD 201091 and HD 95735) the multiple cycles
were detected. Using wavelet analysis the following results (other
than periodograms from Baliunas et al., 1995) were obtained:

HD 131156A shows variability on two time scales: the shorter cycle
is about 5.5-yr, a longer-period variability is about 11 yr.

For HD 131156B only one long-term periodicity has been determined.

For HD 100180 the variable cycle of 13.7-yr appears in the beginning
of the record; the period decreases to 8.6-yr by the end of the
record. The results in the beginning of the dataset are similar to
those found by Baliunas et al. (1995), who found two cycles, which
are equal to 3.56 and 12.9-yr.

The record for HD 201092 also exhibits two activity cycles: one is
equal to 4.7-yr, the other has a time scale of 10-13 years.

The main cycle, seen in the record of HD 201091, has a mean length of
6.7-yr, which slowly changes between 6.2 and 7.2-yr. A shorter,
significant cycle is found in the first half of the record with a
characteristic time scale of 3.6-yr.

The stronger cycle of HD 95735 is 3.9-yr. A longer, 11-yr cycle is also present with a smaller
amplitude.

In our paper we have applied the wavelet analysis for partially available data
from the records of relative CaII emission fluxes - the variation of $S_{HK}$ for
1965-1992 observation sets from Baliunas et al. (1995) and for 1985-2002 observations from
Lockwood et al. (2007). We used the detailed plots of $S_{HK}$ time
dependencies: each point of the record of observations, which we
processed in this paper using wavelet analysis technique,
corresponds to three months averaged values of $S_{HK}$.

In this paper we have
studied 5 HK-project stars with cyclic activity of the "Excellent"
class: HD 10476,  HD 81809, HD 103095, HD 152391, HD 160346 and
the star HD 185144 with no cyclicity.

We used the complex Morlet wavelet 1.5 - 1 which can most accurately
determine the dominant cyclicity as well as its evolution in time in
solar data sets at different wavelengths and spectral intervals
(Bruevich \& Yakunina 2015).

We hope that wavelet analysis can help to study the temporal
evolution of  CA cycles of the stars. Tree-month averaging also
helps us to avoid the modulation of observational $S_{HK}$ data by
star's rotations.

In Figure 6 we present our results for cycles of 6 HK-project stars:

HD 81809 has a mean cycle duration of 8.2-yr, which slowly changes between
8.3-yr in the first half of the record and 8.1-yr in the middle and
the end of the record while Baliunas et al. (1995) found 8.17-yr.

HD 103095 has a mean cycle duration of 7.2-yr, which slowly changes between
7.3-yr in the first half of the record, 7.0-yr in the middle and
7.2-yr in the end of the record while Baliunas et al. (1995) found
7.3-yr.

HD 152391 has a mean cycle duration of 10.8-yr, which slowly changes between
11.0-yr in the first half of the record and 10.0-yr in the end of
the record while Baliunas et al. (1995) found 10.9-yr.

HD 160346 has a mean cycle duration of 7.0-yr which does not change during
the record in agreement with Baliunas et al. (1995) estimated
7.0-yr.

HD 10476 has a mean cycle duration of 10.0-yr in the first half of the
record, then it sharply changes to 14-yr, while Baliunas et
al. (1995) found 9.6-yr. After changing the high amplitude cycle's
period from 10-yr to 14-yr in 1987, the low amplitude cycle remained
with 10.0-yr period -- we can see two activity cycles. Baliunas
et al. (1995) estimated HD 10476 cycle as 9.6-yr.

HD 185144 has a mean cycle duration of 7-yr which changes between 8-yr in
the first half of the record and 6-yr in the end of the record while
Baliunas et al. (1995) haven't found the well-pronounced cycle.

In Olah et al. (2009), the multiple cycles were found for 
HD 13115A, HD 131156B,  HD 93735 stars,
for which no cycles have been found in Baliunas et al. (1995).
For the stars of the "Excellent" class HD 201091 and HD 201092,
cycle periods found in Baliunas et al.
(1995) were confirmed and the shorter cycles (similar to solar
quasi-biennial) were also determined.

Olah et al. (2009) have concluded that all the stars from their
pattern of cool main sequence stars have cycles and most of the
cycle durations change systematically.

However we can see that the stars of the "Excellent"
class have relatively constant cycle durations~-- for these stars the
cycle's periods calculated in Baliunas et al. (1995) and cycle's
periods found with the use of the wavelet analysis are the same.

\begin{figure}[h!!!]
\includegraphics[width=70mm]{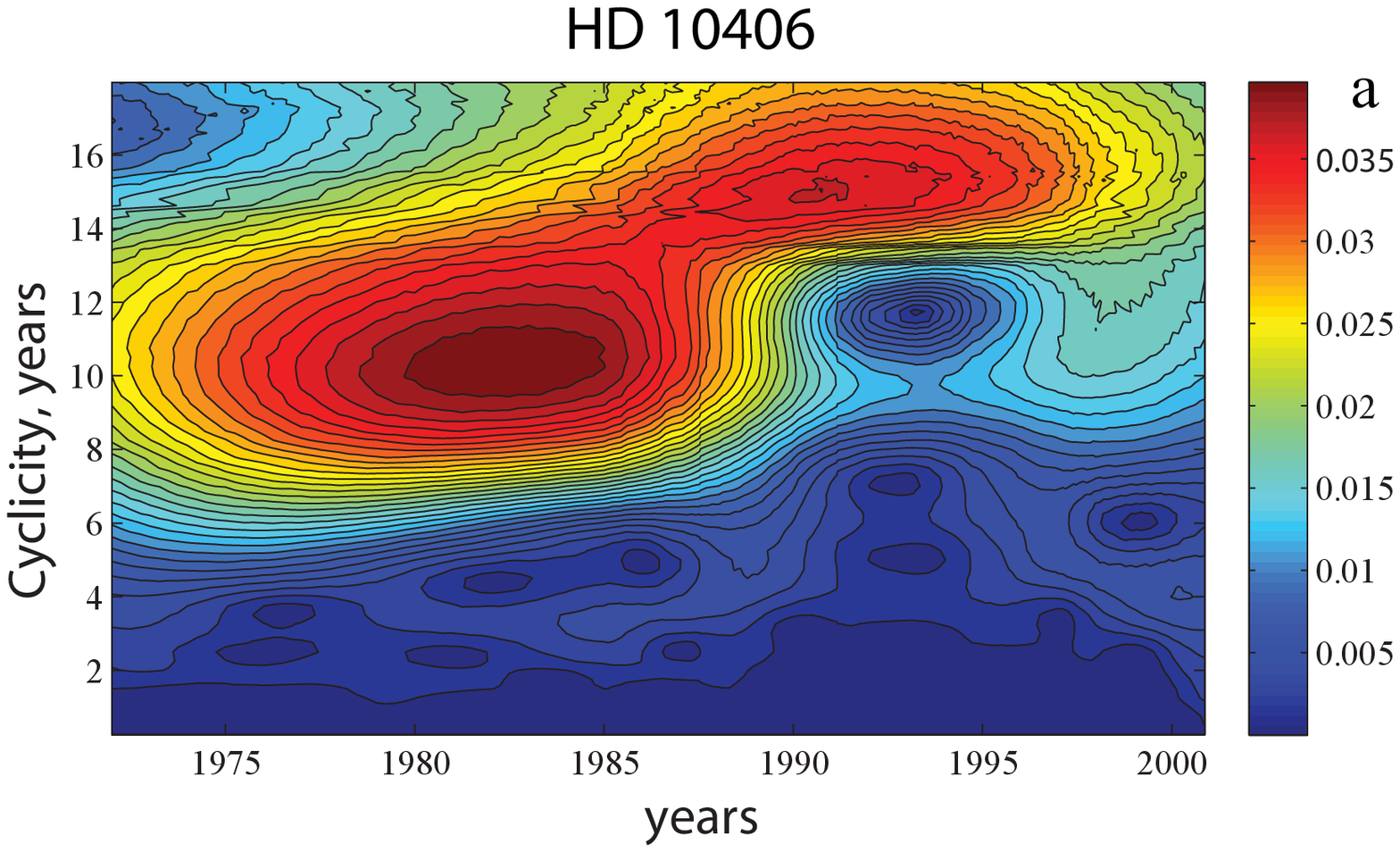}
\includegraphics[width=70mm]{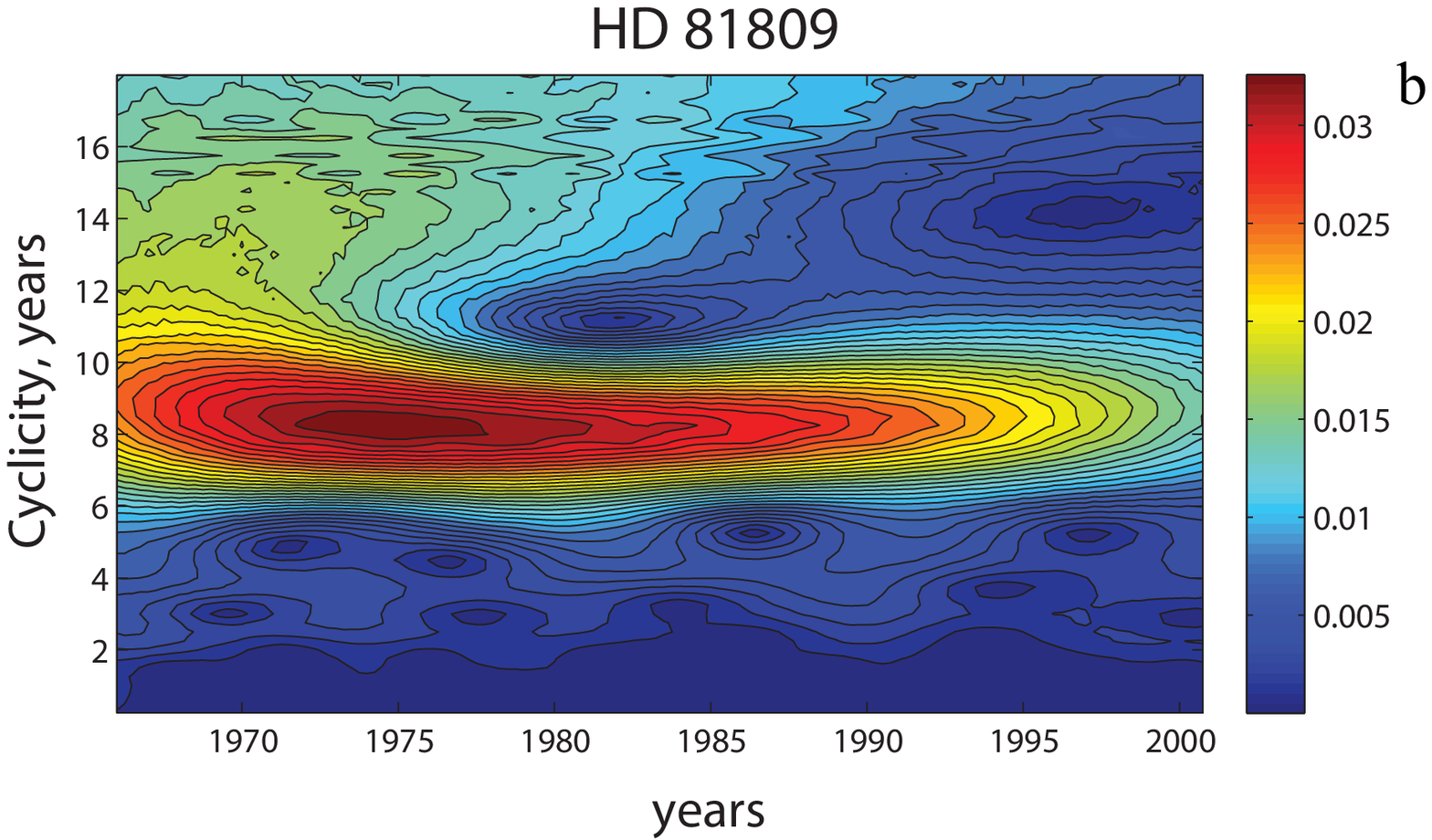}

\includegraphics[width=70mm]{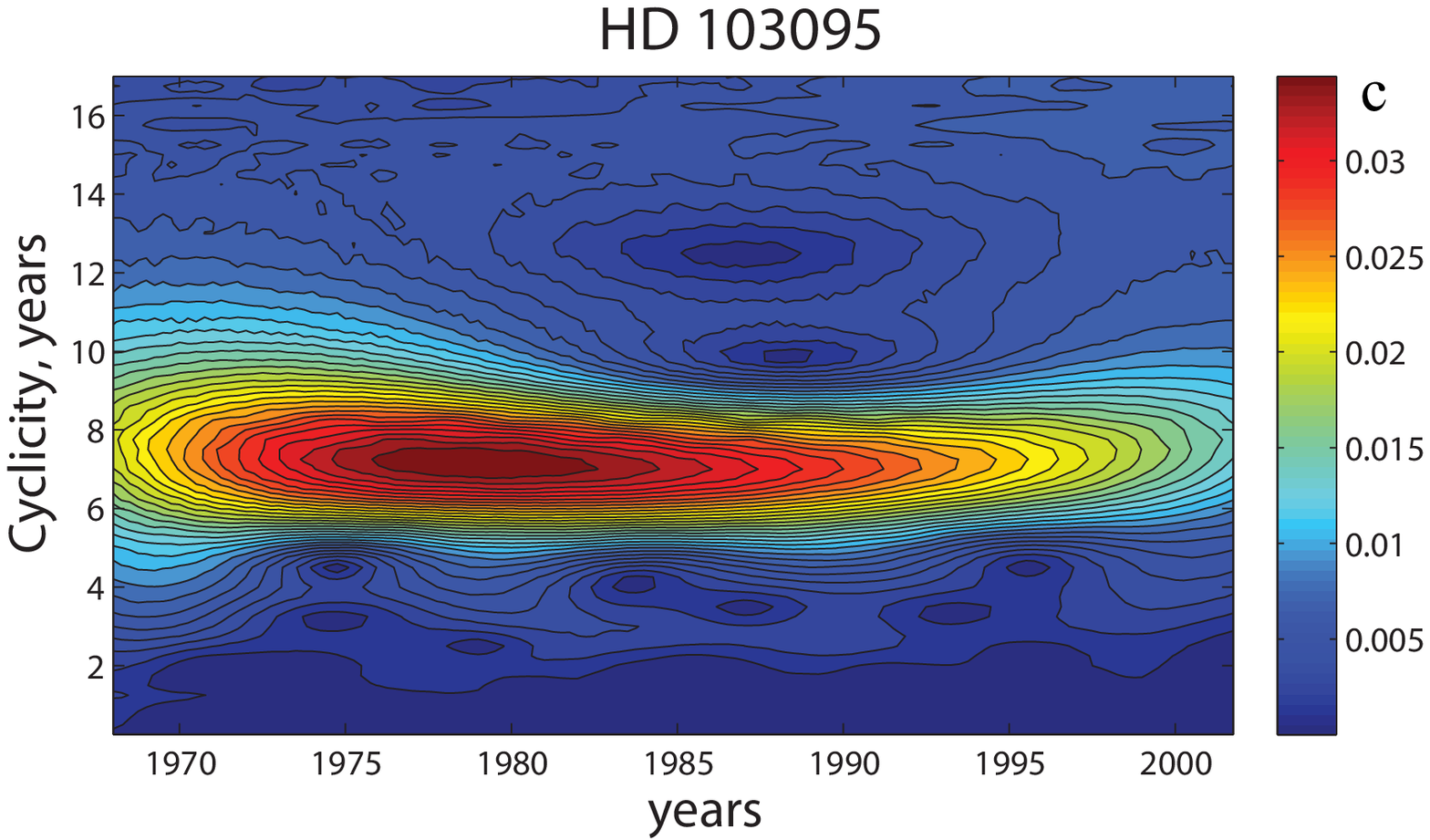}
\includegraphics[width=70mm]{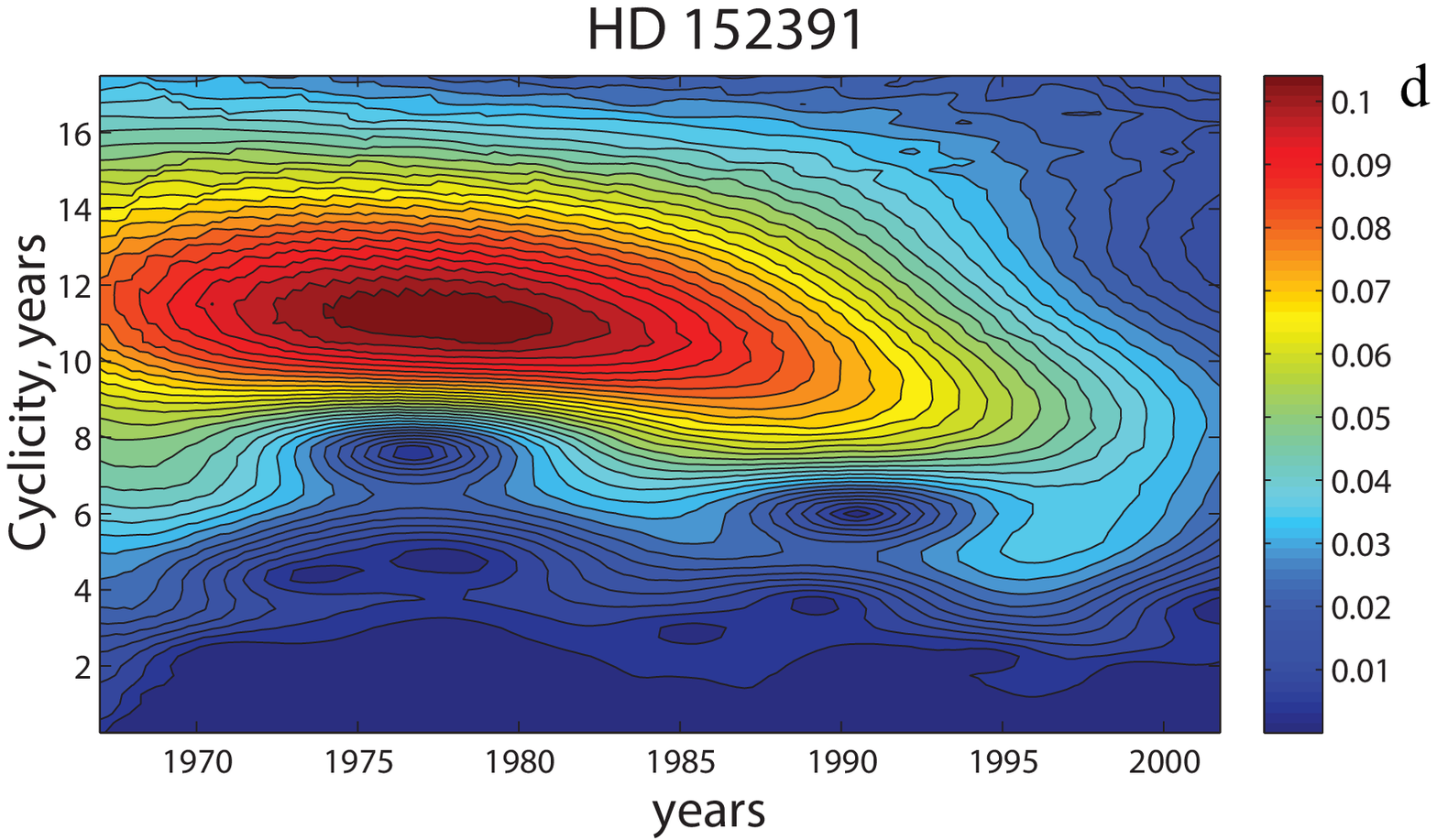}

\includegraphics[width=70mm]{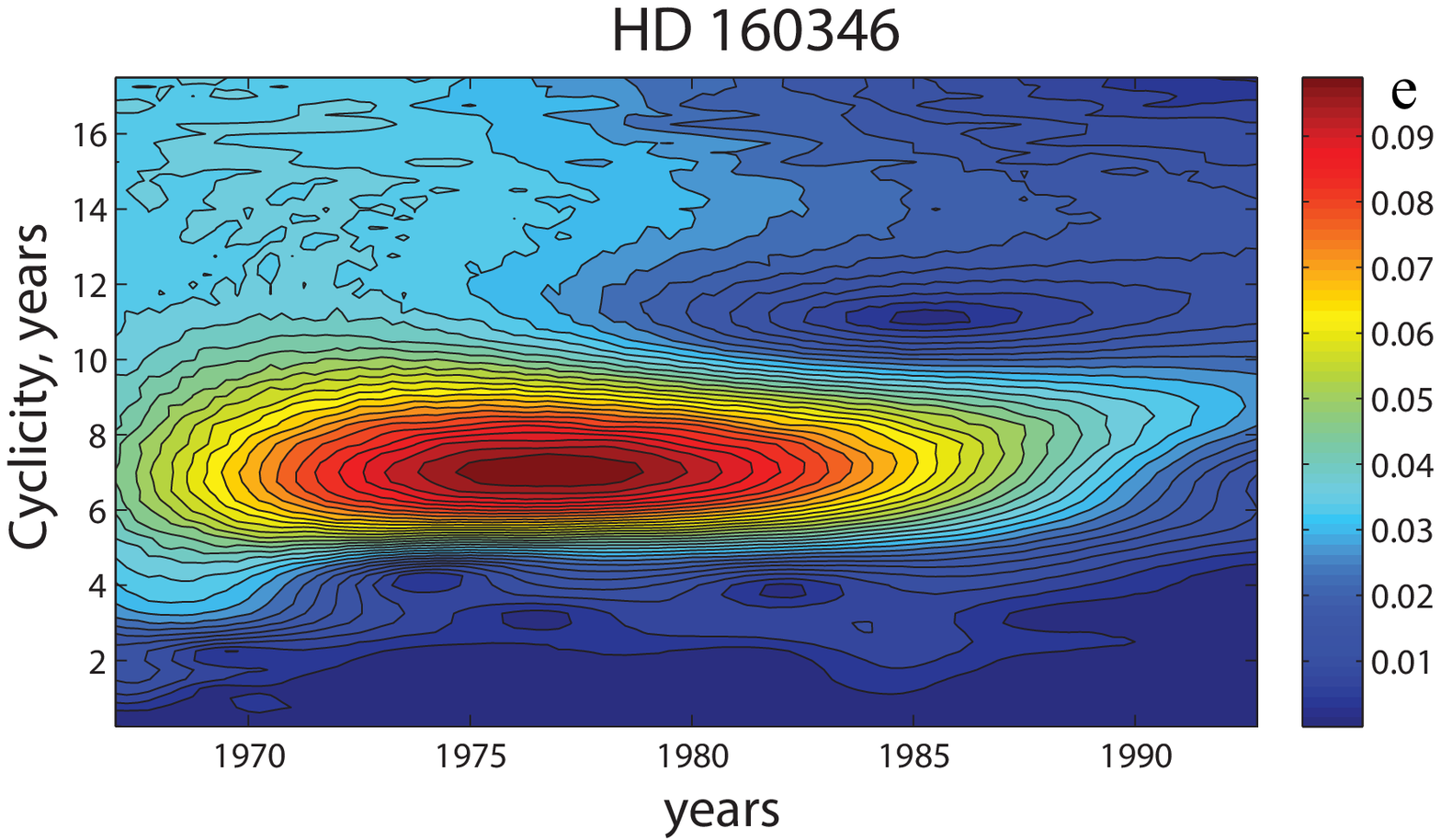}
\includegraphics[width=70mm]{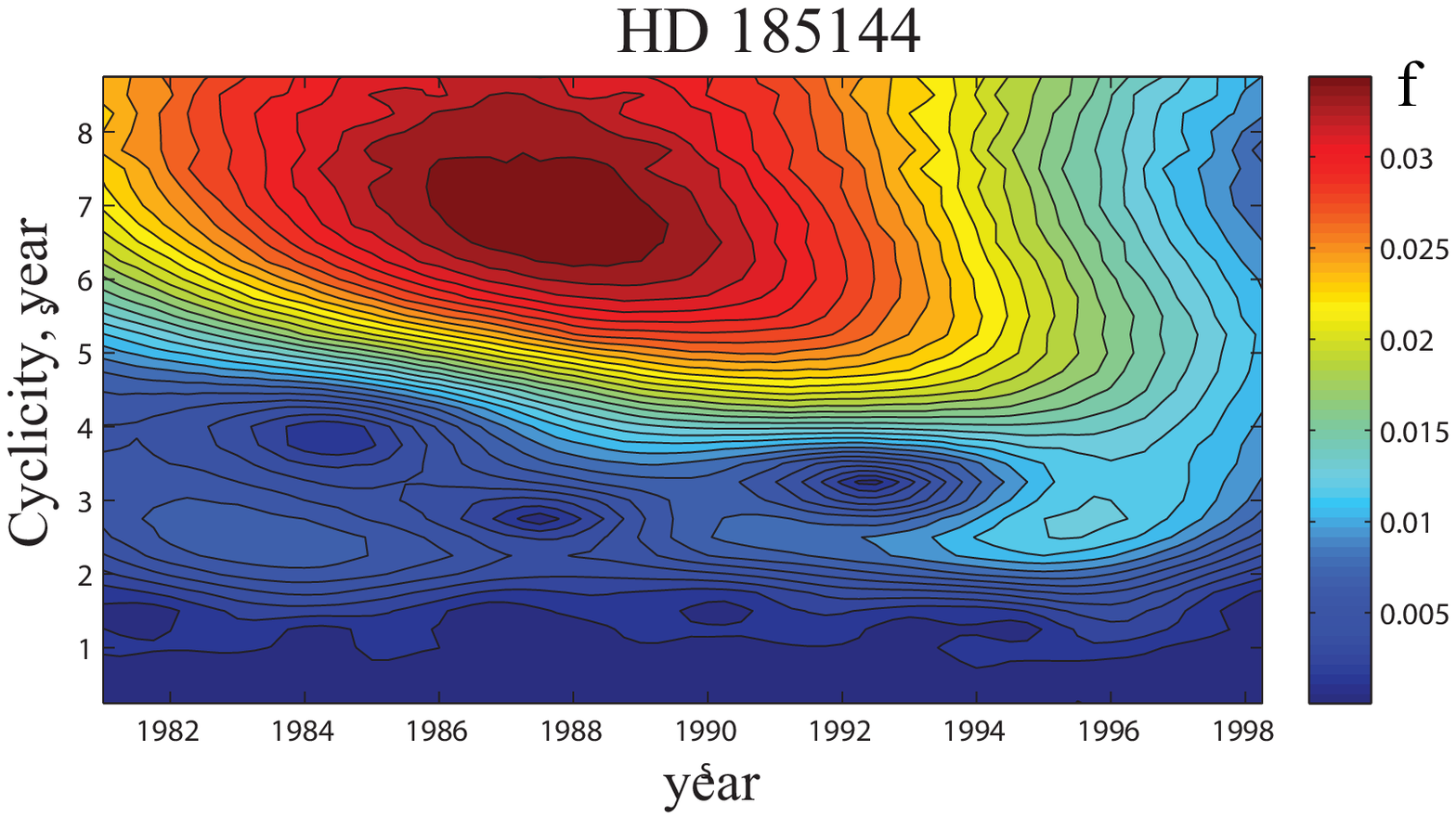}

 \begin{minipage}[]{120mm}
\parbox[t]{120mm}{Fig. 6. Wavelet analysis of HK-project stars. Complex Morlet
wavelet 1.5-1. Observations from 1969 to 2002: (a) HD 10476, (b) HD
81809, (c) HD 103095, (d) HD 152391, (e) HD 160346, (f) HD 185144.}

\end{minipage}

\end{figure}

A similar picture can be seen for the Sun:
the long-term behaviour of the sunspot group numbers has been
analysed using a wavelet technique by Frick et al. (1997) who plotted
changes of the Schwabe cycle (length and strength) and studied
the grand minima. The temporal evolution of the Gleissberg cycle can
also be seen in the time-frequency distribution of the solar data.
According to Frick et al. (1997), the Gleissberg cycle is as variable
as the Schwabe cycle. It has two higher amplitude occurrences: the first
one is around 1800 (during the Dalton minimum), and the next one is around 1950.
They found very interesting fact~-- the continuous decrease in the
frequency (increase of period) of the Gleissberg cycle. While near 1750
the cycle duration was about 50 yr, it lengthened to approximately 130
yr by 1950.

In the late part of the XX century, some of solar physicists began to examine
with different methods the variations of relative sunspot numbers
not only in the high amplitude 11-yr Schwabe cycle but in low amplitude
cycles approximately equal to half (5.5-yr) and fourth
 (quasi-biennial) parts of the period of the main 11-yr cycle, see Vitinsky et al. (1986).
The periods of the quasi-biennial cycles vary considerably within
one 11-yr cycle, decreasing from 3.5 to 2 yrs (see Figure 5b,c),
which complicates a study of such periodicity with the periodogram method.

Using methods of frequency analysis of signals, the
quasi-biennial cycles have been studied not only for the relative
sunspot number, but also for 10.7 cm solar radio emission and for
some other indices of solar activity, see
Bruevich \& Yakunina (2015). It was also shown that the cyclicity
on the quasi-biennial time scale often takes place among stars
with 11-yr cyclicity, see Bruevich \& Kononovich (2011).

The cyclicity similar to the solar quasi-biennial was also detected
for Sun-like stars from direct observations.
In Morgenthaler et al. (2011), results of direct observations of magnetic cycles of
19 Sun-like stars of F, G, K spectral classes within 4 years were
presented. Stars of this sample are characterized by masses
between 0.6 and 1.4 of the solar mass and by rotation periods between 3.4
and 43 days. Observations were made using NARVAL spectropolarimeter
(Pic du Midi, France) between 2007 and 2011. It was shown that for
the stars of this sample $\tau$ Boo and HD 78366 (the same of the
Mount Wilson HK-project) the cycle lengths derived by CA
 by Baliunas et al. (1995) seem to be longer than those
derived by spectropolarimetry observations of Morgenthaler et al.
(2011). They suggest that this apparent discrepancy may be due to
the different temporal sampling inherent to these two approaches, so
that the sampling adopted at Mount Wilson may not be sufficiently
tight to unveil short activity cycles. They hope that future
observations of Pic du Midi stellar sample will allow them to
investigate longer time scales of the stellar magnetic evolution.

For the Sun-like F, G and K stars according to {\it Kepler}
observations, "shorter" chromosphere cycles with periods of about
two years have also been found (see Metcalfe et al. 2010; Garcia
et al. 2010).

We assume that precisely these quasi-biennial cycles were identified
in Morgenthaler et al. (2011): $\tau$ Boo and HD 78366 are the same
of the HK-project, these stars have cycles similar to the
quasi-biennial solar cycles with  periods of a quarter of the
duration of the periods defined in Baliunas et al. (1995).

Note, that in case of the Sun, the amplitude of variations of the
radiation in quasi-biennial cycles is substantially less than the
amplitude  of variations in main 11-yr cycles. We believe that this
fact is also true for all Sun-like stars of the HK-project and in
the same way for $\tau$ Boo and HD 78366.

The quasi-biennial cycles cannot be detected with the Scargle's
periodogram method. But methods of spectropolarimetry from
Morgenthaler et al. (2011) allowed detecting the cycles with 2 and
3-yr periods. Thus, spectropolarimetry is more accurate method for
detection of cycles with different periods and with low amplitudes
of variations.

So, the need for wavelet analysis of HK-project observational data
is dictated also by the fact that the application of the wavelet method
to these observations will help: (1) to find cyclicities with
periods equal to a half and a quarter from the main high amplitude
cyclicity; (2) to clarify periods of high amplitude cycles
and to follow their evolution in time; (3) to find other stars
with cycles for which cycles were not determined using the
periodogram method due to strong variations of the period as in the
case of HD 185144.

The analysis of cyclic activity of Sun-like stars using Scargle's
periodogram method in Baliunas et al. (1995) and wavelet analysis
simultaneously showed that the ranking of stars into classes
according to the quality of their cycles ("Excellent", "Good",
"Fair" and "Poor") is very important moment in the study of stellar
cycles.

Wavelet analysis helped us to understand why stars of "Fair" and
"Poor" classes differ from stars of "Excellent" and "Good"
classes: the main peak on their periodograms is greatly expanded due
to strong variations of the cycle's duration.

As it turned out, the differentiation of stars with cycles into
"Excellent" , "Good", "Fair" and "Poor" classes is very important:
stars with stable cycles ("Excellent" and "Good")
and stars with unstable cycles ("Fair" and "Poor")
relate to different groups in the graphs of dependencies $ log R'_{HK}$ versus
$(B-V)$, log $L_X/L_{Bol}$ versus $(B-V)$,  $P_{cyc}$ versus Age,
see Figure 7 and Figure 8 below.

\vskip12pt
\section{Chromospheric and coronal activity of HK project stars of different spectral classes with cycles}
\vskip12pt

Processes, that determine complex phenomena of stellar activity and
cover practically the whole atmosphere from the photosphere to
the corona, occur differently among Sun-like stars belonging to
different spectral classes.

The Mount Wilson HK project observational data allow us to study the
Sun-like cyclic activity of stars simultaneously with their
chromospheric and coronal activity. The selected data of X-ray
luminosities $log L_X/L_{Bol}$ of 80 HK-project stars were taken from
the ROSAT All-Sky Survey, see Bruevich et al. (2001).

\begin{figure}[h!!!]
\includegraphics[width=70mm]{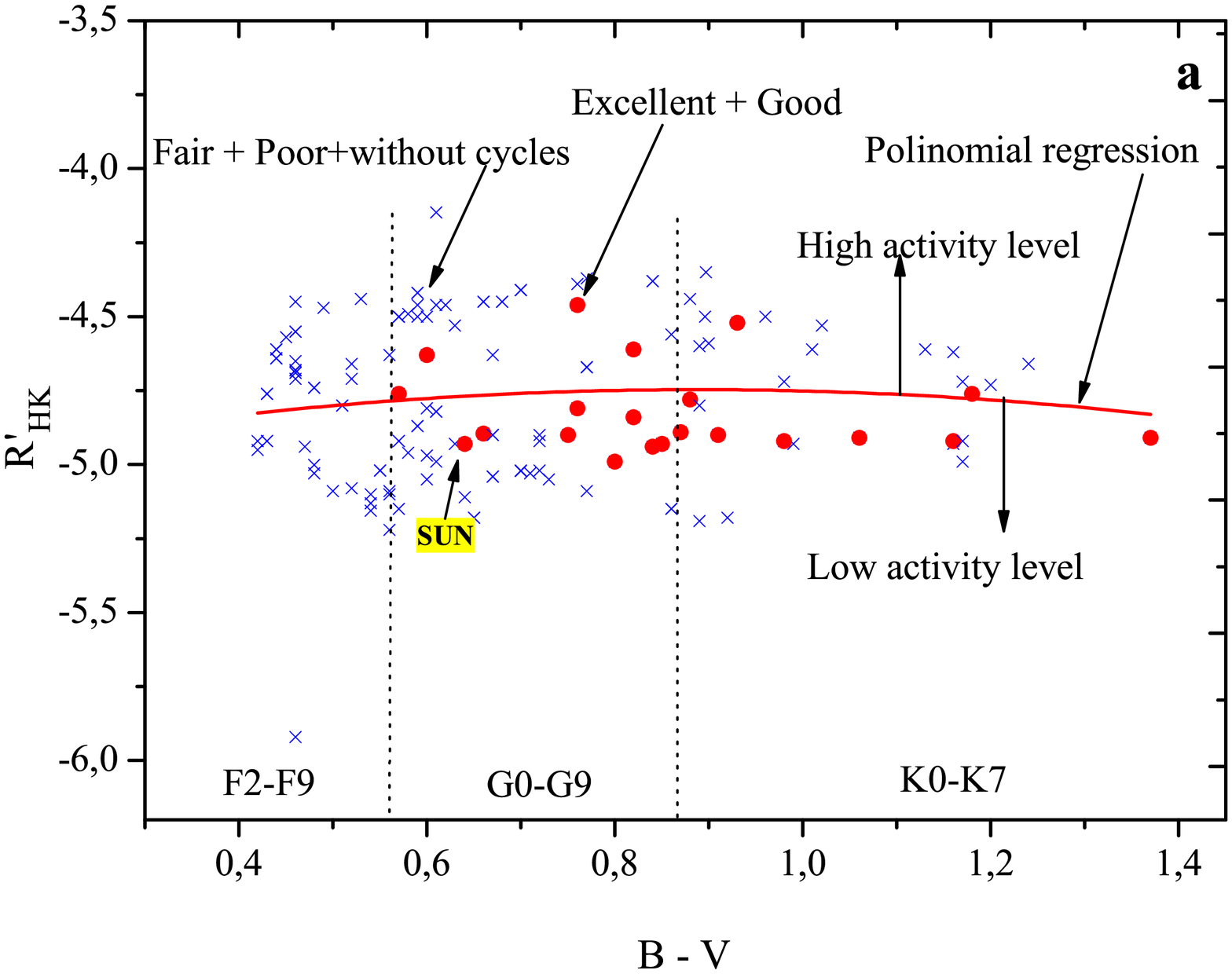}
\includegraphics[width=71mm]{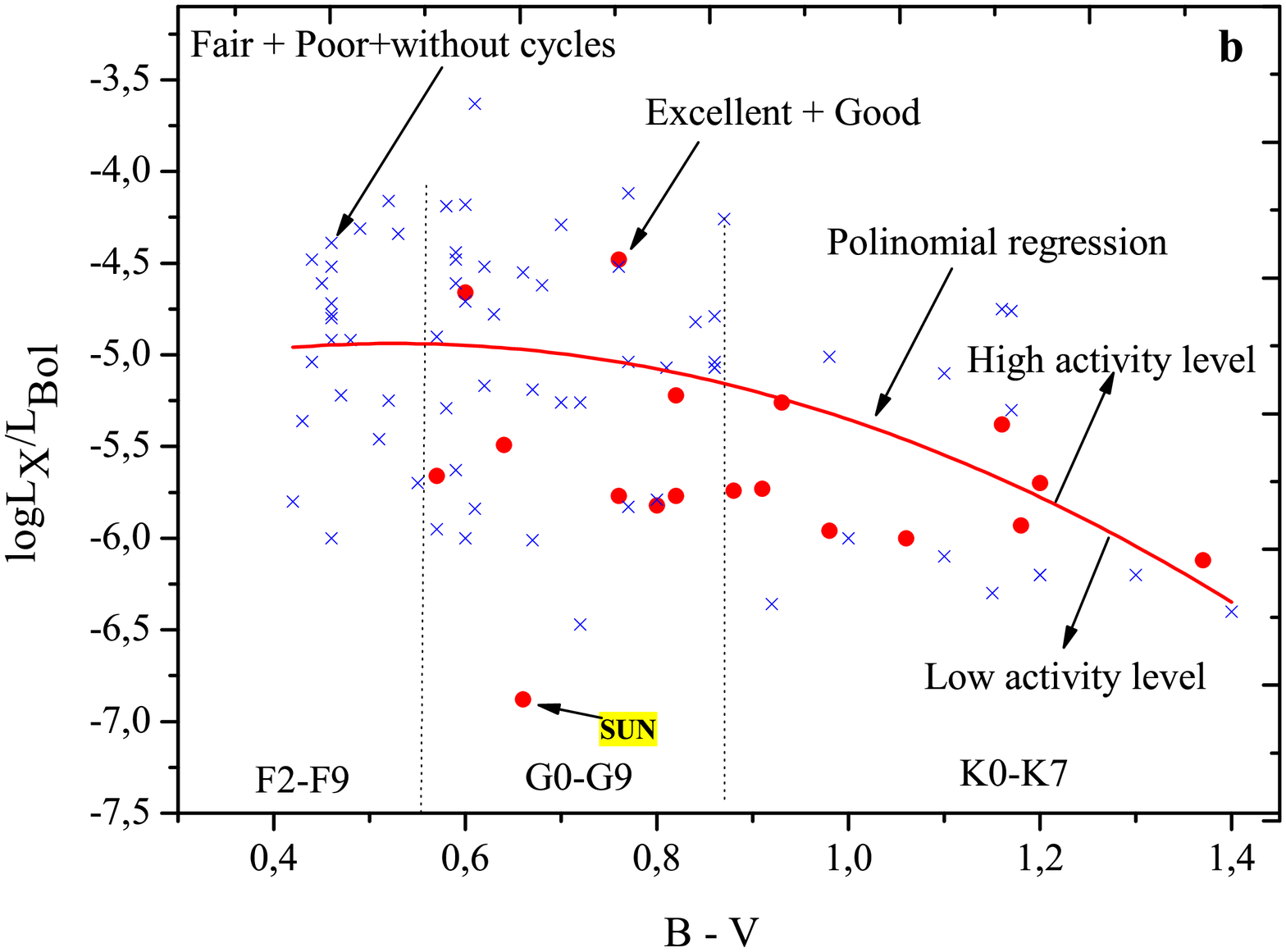}

   \centering
   \includegraphics[width=7.2cm, angle=0]{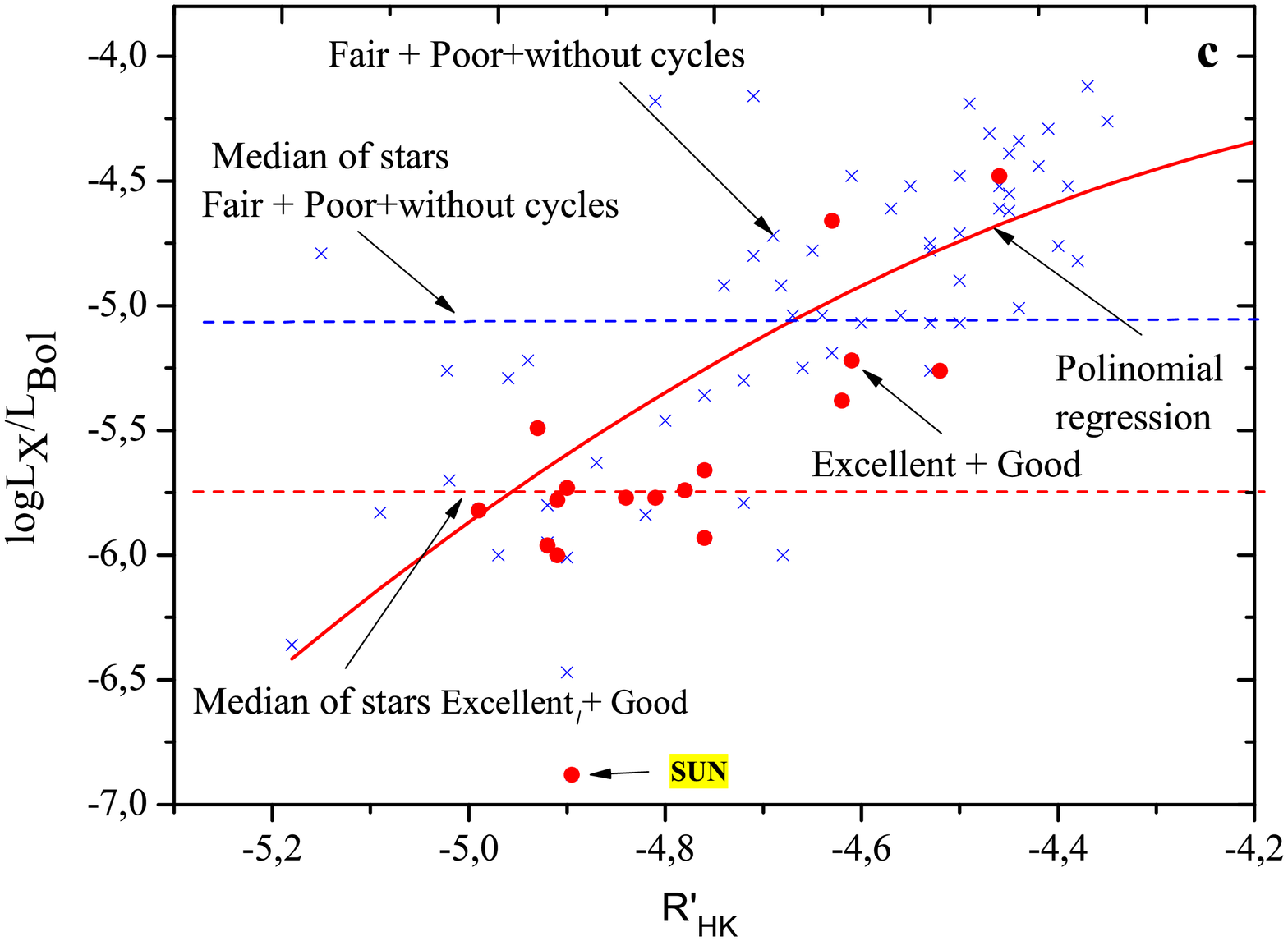}

   \begin{minipage}[]{120mm}
\parbox[t]{120mm}{Fig. 7. HK-project stars. Observations from 1969 to 1994 yrs. (a)
$log R'_{HK}$ versus $(B-V)$, (b) $log L_X/L_{Bol}$ versus $(B-V)$,
(c)  $log L_X/L_{Bol}$ versus $log R'_{HK}$. Solid line~-- polynomial regression,
"Excellent" + "Good" stars~-- filled
circles, "Fair" + "Poor" stars and stars without cycles~-- crosses.
Dashed lines~-- median values for stars of "Fair" + "Poor" +
"Var" classes (top dashed line) and "Excellent" + "Good" classes (lower dashed line).}

\end{minipage}
   \end{figure}

As noted earlier by (Baliunas et al. 1995; Lockwood et al. 2007), the average CA of the
stars, or rather the values of $S_{HK}$ and also of $log R'_{HK}$ varies (increases) with
the increase of the color index $(B-V)$, see Figure 2, Figure 7a.

Our polynomial regression analysis of HK-project stars showed that
there is a relation which is described by the following formula:

    $$  log R'_{HK} = ~-~5.03 + 0.637  \cdot (B-V)  -
  ~ 0.358 \cdot {(B-V)}^2 ~~    (1)  $$

Let us denote the right part of the relation (1) as $F(B-V)$.

We consider the stars which have $log R'_{HK} > F(B-V)$ to be
characterized by the high level of the CA, and stars with $ log R'_{HK}
\leqslant F(B-V)$~-- by the low level of the CA, see Figure 7a.

Next, we have analysed all 110 stars from the HK-project and the Sun
to determine which kind of the level of the CA
corresponds to one or another star. We will consider these results
further in the comparative analysis of stars of different spectral
classes, see Table 1.

For 80 stars, the coronal radiation of which we know from the ROSAT
data, we also do polynomial regression analysis and obtain the
following relationship between the X-ray luminosity, normalized to the
bolometric luminosity, and the color index $(B-V)$:

    $$ log L_X/L_{Bol} = ~-~5.45 + 1.95 \cdot (B-V)
     - ~ 1.85 \cdot {(B-V)^2}   ~ ~  (2)  $$

Let us denote the right hand side of the relation (2) as $P(B-V)$.
By analogy with the analysis of the CA of stars, we consider the
stars with $ log L_X/L_{Bol}  > P(B-V)$ to be characterized by the
high level of the coronal activity, and stars with  $ log L_X/L_{Bol}
\leqslant P(B-V) $~-- by the low level of the coronal activity, see
Figure 7b.

As noted above, in the case of the CA a direct correlation takes place:
with the increase of the color index $(B-V)$ the average value of CA
defined as $S_{HK}$ increases, see Isaacson et al. (2010). When we use
$ log R'_{HK}$ as the index of the CA we see the other trend~-- with
the increase of the color index $(B-V)$ the average value of $  log
R'_{HK}$ is almost constant, see Figure 7a. When we consider the
dependence of the X-ray radiation of stars from $(B-V)$, the inverse
correlation takes place: with the increase of the color index
$(B-V)$, the average value of $  log L_X/L_{Bol}$ decreases, see Figure 7b.

Figures 7a, 7b also demonstrate that stars with cycles of
"Excellent" and "Good" classes are mostly characterized by the low level
of the chromospheric and coronal activity (about 70 \%), as opposed to
stars with cycles of "Fair" and "Poor" classes which are mostly
characterized by the high level of the chromospheric and coronal activity (about 75 \%).

Note that most of stars, characterized by increased CA, have also
increased coronal activity. About 15 \% of stars, including the Sun,
are characterized by the coronal activity that is significantly lower
than the value which should correspond to its CA, see Figure 7c. The
regression line in Figure 7c
divides the stars with relatively high and relatively low $ logL_X/L_{Bol}$. It is seen that the
 stars with no cycles ("Var") and the stars with poorly pronounced cyclical activity ("Fair" + "Poor") are
characterized by relatively high fluxes of coronal radiation (median value corresponds to the
value of $ logL_X/L_{Bol} =-5.05$). The stars, belonging to classes "Excellent"+ "Good", are on
average characterized by lower level fluxes of coronal radiation (a median value corresponds
to the value of $ logL_X/L_{Bol} =-5.75$).

The existence or absence of a pronounced cyclicity, as well as the
quality of the identified cycles (belonging to classes "Excellent",
"Good", "Fair", "Poor"), for F, G and K stars varies significantly, see Table 1.

Bruevich et al. (2001) noted the difference between stars of
"Excellent", "Good", "Fair" and "Poor" classes from a position of
presence and degree of development of under photospheric convective
zones of stars of different color index $(B-V)$.

In Table 1 we can see that for our sample of HK-project stars the
coronal activity is higher in stars of the spectral class F, due to
their total increased atmospheric activity (as compared to stars of
spectral classes G and K). This conclusion is consistent with the
findings of Kunte et al. 1988 in which it was found that
$logL_X/L_{Bol}$ is well correlated (decreases) with the age of stars. But
on the other hand, G and K HK-project stars in our study
are older than F stars that is evident from their slower rotation.

Thus, we can note here (illustrated below in Table 1) that the
quality of CA cycles (the ratio of the total number of stars
belonging to classes with a well-defined cyclicity "Excellent" +
"Good" to the number of stars with less than a certain cyclicity
"Fair" + "Poor") essentially differs for stars of different spectral
classes F, G and K.

\begin{table}
\caption{Comparative analysis of cycles of stars and the quality of
their cyclicities for stars of different spectral classes.}
\centering
\begin{tabular}{clclclclclclclcl}

\hline
                      &      &     &      \\
  Interval of spectral classes &    F2~-F9    &    G0~-G9 ~  &      K0~-K7   \\
\hline
                      &      &     &      \\
 $\Delta (B-V) $ & 0.42~-0.56   &0.57~-0.87~ &0.88~-1.37 \\
\hline
Total number of stars &      &     &      \\
in spectral interval    &39&44  &~27      \\
\hline
   Number of stars with &      &     &      \\
   known values  $L_X/L_{Bol}$ & 26&32   & ~22         \\
\hline
Relative number of stars  with   &      &     &      \\
  increased coronal activity    & 61 \% &   ~42 \% ~  & 36 \%         \\
\hline
 Relative number of stars  with &      &     &      \\
 increased CA   & 56\% &  ~46 \% ~   &   48 \%     \\
\hline
 Relative number of stars with &      &     &      \\
 CA cycles   & 25\% &  ~40\% ~      &   72\%         \\
\hline

Quality of chromospheric cycles   &      &     &      \\
"Excell+Good"/"Fair+Poor"  & 0/10 &  ~7/10 ~      &   14/4         \\
\hline
\end{tabular}

\end{table}

\begin{figure}[h!!!]
   \centering
   \includegraphics[width=10.0cm, angle=0]{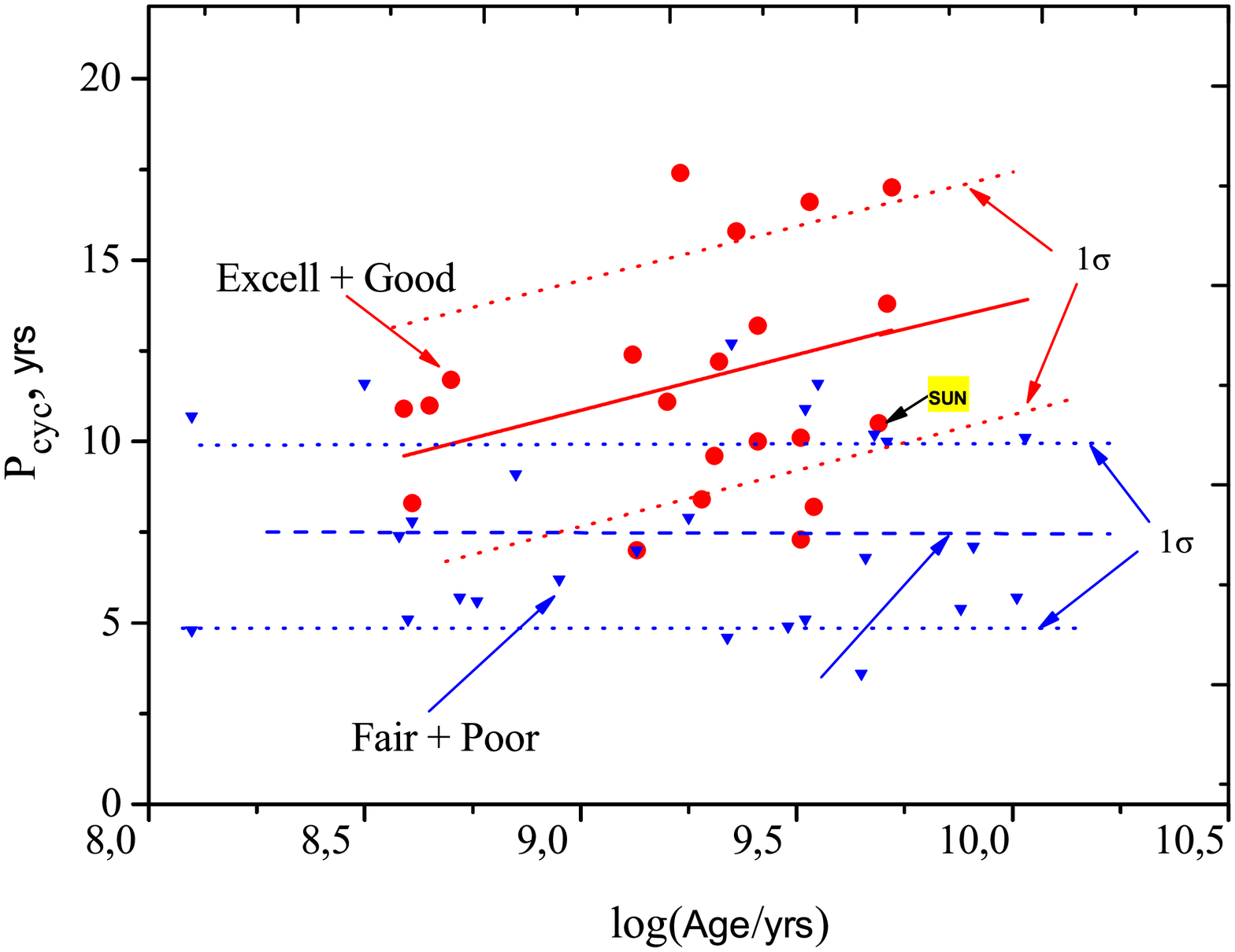}

   \begin{minipage}[]{100mm}
\parbox[t]{120mm}{Fig. 8. $P_{cyc}$ versus stellar age for the stars of the HK-project.
 The regressions for the full sample (the lower dashed line) and for stars with cycles of
 the "Excellent" + "Good" classes (the upper solid line) are shown. For both
 regressions the dotted lines indicate the intervals corresponding to 1$ \sigma$ standard deviation.}

\end{minipage}
   \end{figure}

Different tests of the dependency of the cycle period (with
durations in various time scales from seconds in the asteroseismic
analysis to several yrs in Dynamo processes studies) on different
parameters of Sun-like stars have been performed, see Morgenthaler
et al. (2011); Garcia et al. (2010); Garcia et al. (2014); Mathur et
al. (2012); Metcalfe et al. (2010).

We have analysed the dependence of the star magnetic cycle duration
on their ages. Cycle durations were taken from Baliunas et al. (1995).
Unfortunately, we have very limited stellar sample (Lockwood et al.
2007). Stellar ages were calculated according to Wright et al.
(2004) as a function of the CA.

In Figure 8, the connection of $P_{cyc}$ with ages of stars is
shown. The scatter of points around the regression line is
very large. For stars with cycles of "Excellent" + "Good" classes
with the increase of the age (or various parameters connected with
the age) the duration of cycles increases by about 20 \% with an
increase of log(Age/yr) from 8.5 to 10. The stars with cycles of
"Fair" + "Poor" classes show no dependence of $P_{cyc}$ on age.

For carrying out a high quality examination of cycles for Sun-like stars,
the problem of determination of $P_{cyc}$ as accurately as possible,
using frequency-time (wavelet) analysis, becomes very actual.

\vskip12pt
\section{Conclusions}
\vskip12pt

\begin{itemize}

\item The quality of the cyclic activity, similar to the solar 11-yr one, is
significantly improved (from "Fair + Poor" to "Excellent + Good") in
G and K-stars as compared to F-stars. The F-stars 11-yr cyclicity
(detected only in every fourth case) is determined with a lower
degree of reliability.

\item We show that for the interpretation of observations  the use of modern methods of wavelet analysis is required.
The nature of the cyclic activity of Sun-like
 stars is very similar to the Sun's one: along with main cycles there are quasi-biennial
cycles. Periods of solar quasi-biennial cycles evolve during one
main 11-yr cycle from 2 to 3.5 yrs that complicates their
detection with the periodogram technique. Our conclusion about
quasi-biennial cycles for stars is supported by the direct
observation of cycles with the duration of 2-3 yrs for $\tau$ Boo and
HD 78366 by Morgenthaler et al. (2011) and earlier detection of
cycles with durations of 11.6 and 12.3 yrs by Baliunas et al. (1995) for
the same stars.

\item The level of CA of the Sun is consistent with that of
HK-project stars, which have well-defined cycles of activity ("Excellent + Good") and similar color indexes.

\item The stars belonging to "Excellent"+ "Good" classes are on average
characterized by lower level fluxes of the coronal radiation (a median
value corresponds to the value of $logL_X/L_{Bol}=-5.05$), while
stars without cycles have higher level fluxes of the coronal radiation
(a median value corresponds to the value of $logL_X/L_{Bol}=-5.75$).

\item There is a great interest now to the problem of finding planets of habitable zone
which is the region around a star where a planet with sufficient
atmospheric pressure can maintain liquid water on its surface. We
believe that in search of life on exoplanets the close attention
should be paid to the characteristics of our Sun: a low level of
variability of the photospheric radiation simultaneously with a very
low level of the coronal radiation. So the search for extraterrestrial
life should be conducted simultaneously on the "planets of habitable
zone" and on the "stars comfortable for the life", such as the Sun.

\end{itemize}

\end{document}